\documentclass[sigconf]{acmart}
\pdfoutput=1
\usepackage{booktabs} % For formal tables
\usepackage{subfig}
\usepackage{enumitem,kantlipsum}
\usepackage{multirow}
\usepackage{listings}
\renewcommand\footnotetextcopyrightpermission[1]{} % removes footnote with conference information in first column
\begin{document}
\title{Measuring News Similarity Across Ten U.S. News Sites}
%\titlenote{Produces the permission block, and
%  copyright information}
%\subtitle{Extended Abstract}
%\subtitlenote{The full version of the author's guide is available as
%  \texttt{acmart.pdf} document}

\author{Grant C. Atkins}
\affiliation{%
  \institution{Old Dominion University}
  \city{Norfolk, Virginia}
}
\email{gatkins@cs.odu.edu}

\author{Alexander C. Nwala}
\affiliation{%
  \institution{Old Dominion University}
  \city{Norfolk, Virginia}
}
\email{anwala@cs.odu.edu}

\author{Michele C. Weigle}
\affiliation{%
  \institution{Old Dominion University}
  \city{Norfolk, Virginia}
}
\email{mweigle@cs.odu.edu}

\author{Michael L. Nelson}
\affiliation{%
  \institution{Old Dominion University}
  \city{Norfolk, Virginia}
}
\email{mln@cs.odu.edu}

\renewcommand{\shortauthors}{G. Atkins et al.}

\begin{abstract}

News websites make editorial decisions about what stories to include on their website homepages and what stories to emphasize (e.g., large font size for main story). 
The emphasized stories on a news website are often highly similar to many other news websites (e.g, a terrorist event story). 
The selective emphasis of a top news story and the similarity of news across different news organizations are well-known phenomena but not well-measured. 
We provide a method for identifying the top news story for a select set of U.S.-based news websites and then quantify the similarity across them. 
To achieve this, we first developed a headline and link extractor that parses select websites, and then examined ten United States based news website homepages during a three month period, November 2016 to January 2017.
Using archived copies, retrieved from the Internet Archive (IA), we discuss the methods and difficulties for parsing these websites, and how events such as a presidential election can lead news websites to alter their document representation just for these events. 
We use our parser to extract $k = {1, 3, 10}$ maximum number of stories for each news site. 
Second, we used the cosine similarity measure to calculate news similarity at 8PM Eastern Time for each day in the three months. 
The similarity scores show a buildup (0.335) before Election Day, with a declining value (0.328) on Election Day, and an increase (0.354) after Election Day. 
Our method shows that we can effectively identity top stories and quantify news similarity.

%\footnote{This is an abstract footnote}
\end{abstract}

%\begin{CCSXML}
%<ccs2012>
%<concept>
%<concept_id>10002951.10003227.10003392</concept_id>
%<concept_desc>Information systems~Digital libraries and archives</concept_desc>
%<concept_significance>500</concept_significance>
%</concept>
%</ccs2012>
%\end{CCSXML}

%\ccsdesc[500]{Information systems~Digital libraries and archives}

%\keywords{Web Archiving, Document Representation, Similarity}

\maketitle

%
% Start content
%

\section{Introduction \& Motivation}

We are interested in mining archived news websites, specifically to measure similarity of news across different sites.
This involves identifying what is considered news, how it is represented, and emphasized on a website. 
For example, on January 4, 2018 \textit{foxnews.com}'s coverage of the \textit{Fire and Fury} book \cite{fireFuryBook} was not apparent on their website\footnote{http://web.archive.org/web/20180104210001/http://www.foxnews.com/}; it was mid-way down the page with the headline ``Trump demands publisher halt release of book that led to Bannon fallout.'' 
Their top story was ``Freedom to drill: Trump dramatically expands offshore drilling, opens nearly all US coastal waters to explore oil and gas''.  
At roughly the same time, \textit{msnbc.com}'s top story was \textit{Fire and Fury}, with the top story entitled ``Trump: Bannon changed tune, called me a great man last night'' and there were no stories about offshore drilling\footnote{http://web.archive.org/web/20180104203948/http://www.msnbc.com/} "above the fold" (i.e., viewable in the top portion of the webpage without scrolling).

%There may be days in a week where there is no event of significance and every news site reports on topics different from other sites.
%This would signify that there is a low similarity among all news sites.

Most of the journalistic work that chronicles important events is represented in digital form as online news content. 
Consequently, there has been an increased interest in preserving online news. For example, the ``Dodging the Memory Hole'' initiative \cite{dodgingMemoryHole} emphasizes the importance of long-term preservation of online news content. 
We consider the preservation of online news important as well as the provision of tools\footnote{http://www.pastpages.org/} to study and perform analysis on archived news content.
News organizations, like \textit{The New York Times}, which recently started their own archiving initiatives \cite{shanwangny}, also consider the preservation of old pages to be of importance. 
As described by Hansen \& Paul \cite{hansen2017futureproofing}, many news websites, like \textit{The Wall Street Journal}, do not have screen captures of when their websites officially launched.
Web archives have become a key tool to go back in time and replay these websites.

Web archives can only offer what they have, requiring new tools \cite{archivenow2018} to preserve online content in multiple archives. 
Comparing archived news in aggregate also requires a comparison of their preserved date time.
It is not expected for web archives to have continuous minute-by-minute preservations of web pages which is a tradeoff when compared to the live web.
%When choosing which time to select for measuring similarity we select a specified time when news is archived most frequently, described in detail in Section \ref{memento_sel}. 

To measure similarity in news websites, we retrieve the mementos (i.e., archived web pages) for the months of November 2016, December 2016, and January 2017 from the top-level page of 10 national news websites based in the United States.  
We extracted the headlines and URIs for the top-$k$ (where $k = {1, 3, 10}$) news stories as close as possible to 8PM Eastern time for all 10 sites.

We propose a new tool to parse select news sites HTML documents using CSS selectors to identify top stories and other top headlines.
We considered RSS feeds for this task as they may offer top news stories, however, they are often provided in the order they are created, which may not reflect how an actual homepage for a news site might look.
RSS feeds are also not always reliably or as frequently archived as news homepages.

Our tool aims to solve this issue by identifying the top stories for a specified news site.
Additionally, our tool identifies "Hero stories," which are the most prominent top stories on news sites often emphasized with large font and central placement.
A Hero story, where $k = 1$, is identified by position, font size, and image size (if one exists), depending on the layout of the news homepage. 
If our parser failed to extract the top news story we considered the next subsequent headline parsed as the Hero story.

%We explore the use of cosine similarity scoring to score a collection of news articles, described in Section \ref{sim_metrics}.
We explore the use of the cosine similarity measure to calculate the similarity of a collection.
Over a three month period we show that the count of stories used in the news similarity calculation directly influences the overall similarity scores for a given day.
We show that using these similarity scores we can effectively identify events such as the 2016 presidential election, a national holiday like Thanksgiving, and other significant political events like the announcement of the Travel Ban \cite{execorder2017} (Executive Order 13769).

% Show actual results
Using the cosine similarity scores we identify the Travel Ban event to have the highest overall similarity score regardless of the value of $k$.
We show in Section \ref{topksect} that our similarity score can be used to identify overlapping stories for a given day outside of the election.
We identify that similarity values peak after significant events start.
For example, after Election Day and the Travel Ban announcement the cosine similarity scores increase, indicating that there is a delay in synchronization of news.
% FIX THIS BELOW
Our similarity scores show that we are able to identify a decline, from 0.417 to 0.343, in similarity after the U.S. election period has passed, which indicates news sites pursuing other stories.

\section{Related Work}

%There have been many advances in topic detection and news relevance.
%Lee et al. introduced several approaches for identifying news stories in blogs using prior news headlines as well creating criteria for measuring the importance of a news headline \cite{Lee:2010}.
%We analyze a collection as a whole but determining the importance or influence of an article on a collection is something highly considered when doing pairwise comparisons in a collection.

There are many existing efforts already available in topic detection, news parsing, and building collections. 
Topic Detection and Tracking (TDT) \cite{allan1998topic}, a DARPA-sponsored initiative, introduced and formalized the problem of determining if two new stories are about the same topic or highly similar. 
The ability to determine similarity enables many tasks such as clustering news. 
Lau et al. \cite{lau2012line} developed a method to track emerging events on Twitter using a topic model based on time slices and a dynamic vocabulary. 
He et al. \cite{he2010topic} modeled the problem of detecting bursts in news cycles by using physics concepts such as mass and velocity. 
They showed this approach was effective at accurately and efficiently detecting bursts for MeSH (Medical Subject Heading) terms. 
Similar to these approaches, we modeled news reporting of the November 2016 U.S. election cycle and showed the similarity measure we defined is effective in detecting news similarity and bursts of news synchronization.

We used archived webpages, or mementos, from the Internet Archive in our analysis. 
As demonstrated by Brunelle et al. \cite{brunelle2015not} and Berlin \cite{Berlin:cnn}, not all mementos render correctly upon playback. This affects our method to determine the top stories on a news website.  
In Section \ref{el_day_influences}, we discuss the representations of some of the news websites whose mementos from November 2016 are not rendered consistently.

Klein and Broadwell \cite{Klein:2015} presented analyses on both television news and social media collections showing spikes in attention for continuous news stories and the quick drops that follow afterwards.

Other efforts related to news parsing can be seen in studies conducted by various news organizations. 
The \emph{New York Times} conducted a study \cite{nytimesStudy} that showed how three different news organizations (\textit{Fox News}, \textit{msnbc.com}, and \textit{CNN}) covered the indictment of Paul Manafort and Rick Gates by the special counsel Robert Mueller. 
They showed that \textit{msnbc.com} and \textit{CNN} devoted more airtime to the coverage of this news. 
Similarly, \textit{FiveThirtyEight} \cite{fivethirtyEight} showed that the ideological leanings affected when/what news networks covered about the ``Trump-Russia'' story. 
Even though our research identifies common interests in news reports during the 2016 U.S. election cycle, we do not infer the reasons behind why the news organizations cover various topics. 
However, our analysis could help a domain expert conduct such a study.

Similarity in news has also been studied in the context of identifying media manipulation on social media. 
Wolley et al. \cite{woolley2017computational} studied propaganda bot activities during the 2016 U.S. general election and showed how these bots propagate similar content to boost the prominence of a political candidate in an attempt to ``manufacture consensus''. 
Roger Sollenberger \cite{Sollenberger} showed a similar activity among fringe media organizations by illustrating how they publish similar content and interlink their web pages in an attempt to increase their search engine rankings. 
Similarly, Faris et al. \cite{faris2017partisanship} studied the media landscape of the 2016 U.S. general election in order to identify how partisan news organizations covered Donald Trump and Hillary Clinton. 
They showed that the media overwhelmingly covered Trump more than Clinton. 
These research efforts show that a high degree of similarity between news content is not always due to an organic increase in interest surrounding a topic, but it can be manufactured. 
Our research identifies similarity over time, and since we control the news sources we sample URIs from, we do not focus on assessing if any high degree of similarity in news content is inorganic.

Alvin Chang \cite{voxFoxNewsVis} curated a list of prominent news topics and then compared coverage by \textit{MSNBC}, \textit{CNN}, and \textit{Fox News}, showing \textit{Fox News} has drastically different coverage for these topics.
Our research shows how top stories may be synchronous across many websites but others may choose not to report on it or position it further down the page. 

Many other research efforts that combine topic detection and news parsing are tied toward collection building. 
Nwala et al. \cite{nwala2017local} introduced the Local Memory Project, which provided a suite of tools for archiving, building, and exploring collections from local news stories. 
Hamborg et al. \cite{matrixBased} introduced NewsBird, a news aggregation system that attempts to limit bias in news collections by balancing multiple different perspectives on international news topics. 
Other efforts related to collection building use focused crawlers in order to build collections about specific topics. 
At the center of focused crawling is the means to determine if an incoming URI belongs in a collection. 
Some systems \cite{bergmark2002collection, chakrabarti1999focused} used classifiers to determine if a URI belongs in a collection, others \cite{farag2017focused} have used a similarity measure. 
Even though this research is not focused on collection building, it is relevant to collection building efforts since the news similarity metric can serve as a filter in the extraction of URIs that share a common topic. 

\section{Methodology}

Selecting news homepages outside of a relatively close timeframe can greatly alter the evaluation of the similarity. 
Suppose a user reads the homepage of news website in the morning, do they expect to see the same homepage by the end of the day?
Although it is important that each of these news websites are archived frequently to catch the fast changing news, for our purpose we needed to establish a time from which to collect a single memento where each news homepage was archived and then evaluate upon the set of mementos collected at this time.
This section is organized as follows. First, we explain why our methodology is only possible due to the preservation of news sites in web archives.
Second, we provide an explanation on the chosen news websites and memento collection time. 
Third, we explain the process to extract the top story and subsequent headline titles and links from each of these homepages.
Finally, we describe the similarity measure used to calculate the similarity for a given day over the course of the three selected months.

\subsection{Mining Archived News}

There is an increased interest in archiving news articles and homepages.
When these news pages become archived, they can be used for more than just replay.
For example, archived news pages can be used to study document design, significant events, political biases, editorial decisions, video playback, etc.

The methodology proposed in the following sections is not practical for live webpages mainly due to the following reasons:
\begin{itemize}
\item Parsing news homepages based on CSS selectors requires document representations to be static - unchanging over time. 
We can build static CSS parsers because we can study archived copies for these pages.
This is not possible on the live web due to the fact that sites often change their document representations or CSS naming conventions, which will result in the parsing of zero stories for future time periods.
\item Live web pages can introduce noise into similarity calculations.
In Section \ref{memento_sel} we discuss the reasons why we chose not to select specific news sites due to paywalls.
Duplicate pages such as paywalls \cite{paywallAtkins}, login pages, or empty documents, introduce noise when calculating similarity between documents which is not only limited to news content.
Although archives may also contain the aforementioned pages, it becomes a simpler task using archives to find such pages whereas on the live web it is difficult to differentiate between a news article or a news paywall page.
\end{itemize} 

The challenges mentioned in this section and the document representation changes mentioned in Section \ref{el_day_influences} show that web archives are a key tool in calculating similarity for collections of news pages. 
The use of preserved web pages allows the collection and analysis of news webpages in a controlled way.

\subsection{Memento selection} \label{memento_sel}

Table \ref{tab:news_sites_mcounts} outlines the ten U.S. based newspaper and TV organizations we considered for our analysis.

\begin{table}
\caption{Memento counts for news site homepages for November 2016, December 2016, and January 2017.}
\label{tab:news_sites_mcounts}
\begin{minipage}{\columnwidth}
\begin{center}
\begin{tabular}{l | l | l | l |}
\toprule
\textbf{Site} & \textbf{Nov. 2016} & \textbf{Dec. 2016} & \textbf{Jan. 2017} \\
\midrule
 washingtonpost.com & 3560 & 3950 & 3836 \\
 foxnews.com & 1250 & 1459 & 1313\\
 abcnews.go.com & 708 & 874 & 1193\\
 nytimes.com & 3177 & 3936 & 3809\\
 usatoday.com & 1405 & 1773 & 1546\\
 cbsnews.com & 816 & 741 & 805\\
 chicagotribune.com & 470 & 609 & 565\\
 nbcnews.com & 818 & 1164 & 1048\\
 latimes.com & 950 & 1188 & 989\\
 npr.org & 2208 & 3016 & 2371\\
\bottomrule
\end{tabular}
\end{center}
%\bigskip
%\footnotesize\emph{Source:} This is a table
%sourcenote. This is a table sourcenote. This is a table
%sourcenote.
%\emph{Note:} This is a table footnote.
\end{minipage}
\end{table}

The \textit{Wall Street Journal} (\textit{WSJ}) was also considered for this project, however we discovered a majority of its stories are behind a paywall requiring users to subscribe to their website to view more than a snippet for a story.
Investigations into archived paywalls \cite{archivePaywalls, paywallAtkins} show that news sites are increasingly putting content behind paywalls.
Therefore, we excluded \textit{WSJ} and other sites that only provided snippet stories.

Another site considered for this project was \textit{msnbc.com}. 
Although the archival rate for this site in November 2016 was high, most of the stories shown on the homepage were actually multimedia and would lead to small snippet summaries or no summary.
This website's \textbf{class} attribute naming conventions did not differentiate between the textual stories and the multimedia stories and may cause our parser to introduce false positives.
Therefore \textit{msnbc.com} was also excluded.

We omitted \textit{CNN} because of known issues \cite{Berlin:cnn} with replaying their mementos from the Internet Archive. 
Unfortunately, these issues began in November 2016, the start of our target timeframe.

News sites like \textit{CNN}, \textit{WSJ}, and \textit{msnbc.com}, are not the only sites that have mementos that are afflicted with problems.
When archived sites render their content, stylesheets and JavaScript dependencies may be missing, making it difficult to determine the order and importance of content due to the lack of structure incurred by the missing dependencies.
News sites with paywalled content may resolve \cite{paywallAtkins} in time to actual content in the Internet Archive, but this could also lead to the selection of a different page at another point in time. 
These are a few of the hazards of working with and relying on the content of web archives.

We used the Internet Archive's Memento API to collect the TimeMaps (TM), which are lists consisting of the URIs for archived copies (mementos) of webpages, for each of the news websites in Table \ref{tab:news_sites_mcounts}.
From these TMs we selected all of the mementos from November 2016 to January 2017.
We selected this date range because we acknowledged that a significant event, the United States presidential election, occurs along with three other national holidays: Veterans Day (November 11), Thanksgiving (November 24), and Christmas (December 25), which provided the opportunity to identify these events.

\begin{figure}
\includegraphics[height=2.5in, width=3.3in]{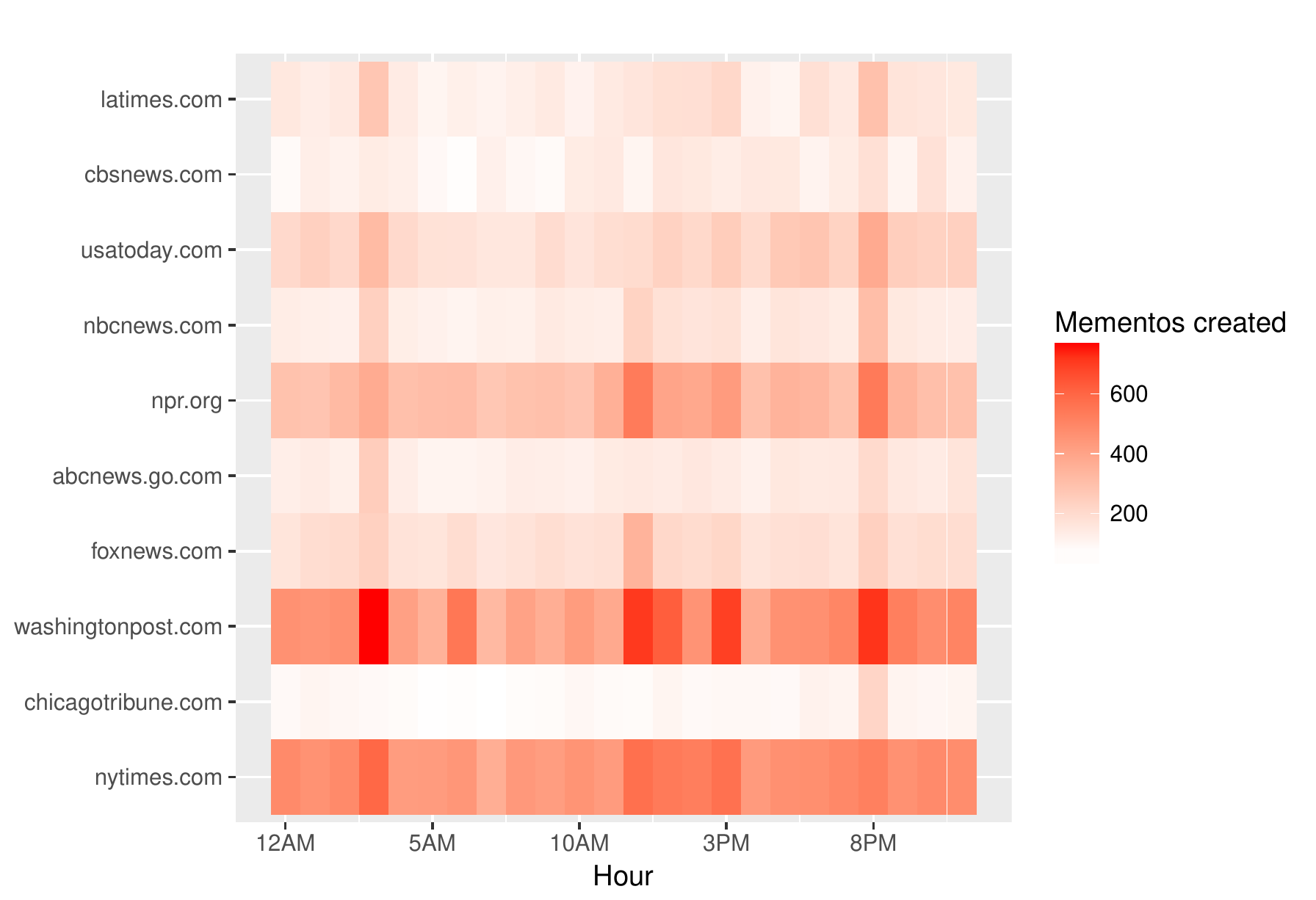}
\caption{Memento creation times by hour, Eastern Time, for the 3 selected months. This shows that 8PM has the highest archival rate.}
\label{fig:heatmap}
\end{figure}

When looking at the archival time for the mementos for the month of November, we found that most of the websites were archived at approximately 1AM GMT, or 8PM Eastern Time, as shown in Figure \ref{fig:heatmap}. 
For the three months, 8 PM Eastern Time had 335 mementos archived at this time while 7 AM Eastern Time had lowest average number of archived mementos at 154 mementos.
We therefore chose 8 PM Eastern Time to ensure we had a higher chance to obtain mementos closer to the same archived time.
We subsequently collected the mementos from the TM at 1AM GMT (8 PM Eastern Time) from 2016-11-01 to 2017-01-31 using the Internet Archive's Memento API.

\begin{figure}
\includegraphics[height=2.8in, width=3.3in]{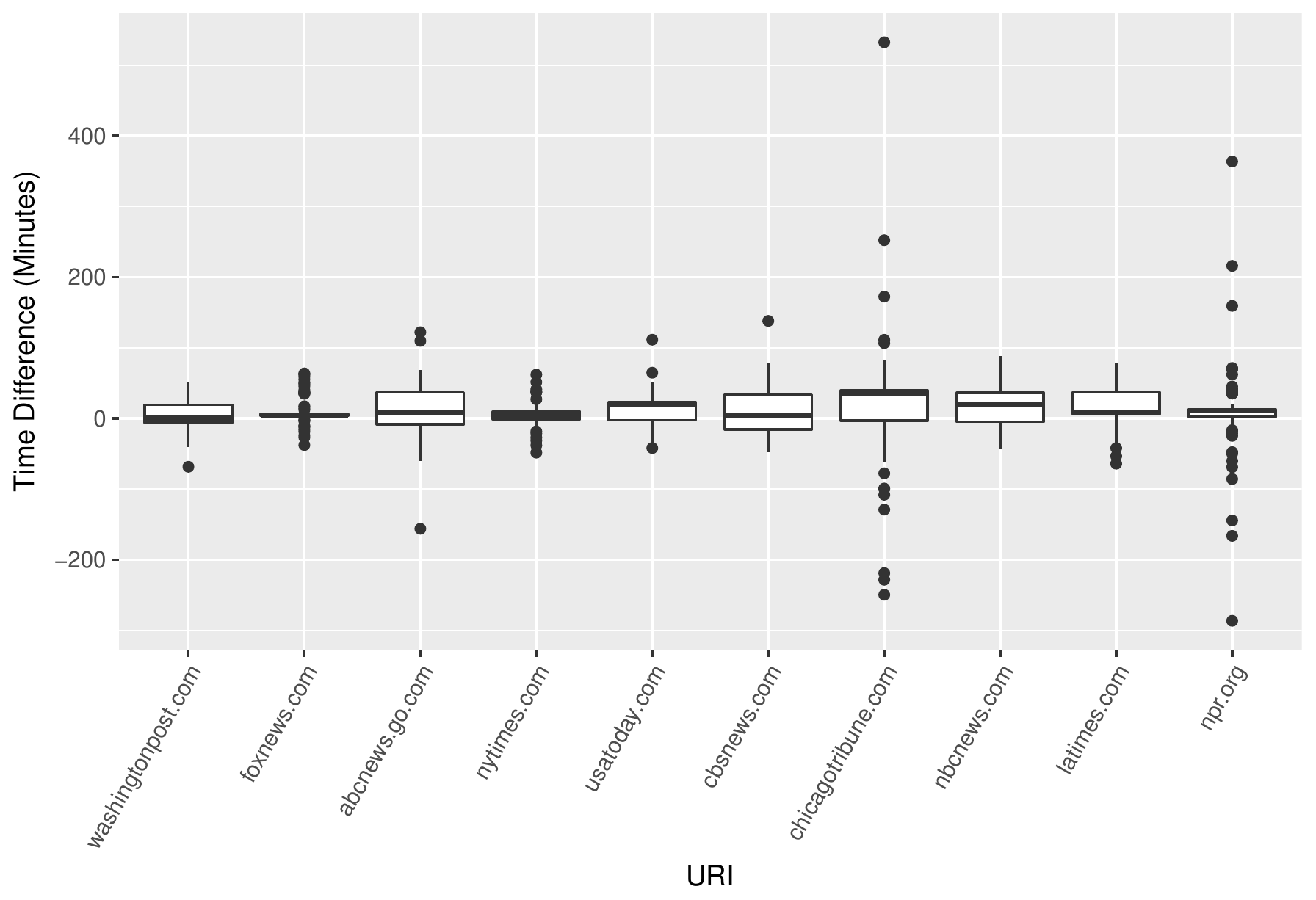}
\caption{Time difference of Mementos collected for each news website from the request time of 8 PM Eastern Time}
\label{fig:boxplot}
\end{figure}

We found that the distribution of minutes from the requested memento time is on average within 100 minutes of the original request, as shown in Figure \ref{fig:boxplot}.
Figure \ref{fig:boxplot} shows the distribution of the distance between the datetime of the URI-M (Memento URI) and the requested time of 8 PM Eastern Time in minutes.
The boxes for each news site represent the overall distribution between their mementos.
A smaller box range shows that a large amount of mementos were retrieved within a close time range.
For example, \textit{foxnews.com} has the smallest box in Figure \ref{fig:boxplot} showing that most of the mementos in the Internet Archive were archived at almost the same time every day.
While \textit{npr.org} also has a small box, there is a greater number of outlier mementos, denoted by dots, that were not tightly archived around the majority time of 8 PM Eastern Time.

Figure \ref{fig:boxplot} shows that \textit{chicagotribune.com} and \textit{npr.org} have multiple mementos beyond the 100 minute average range, showing that these websites may not be as popular for news and are therefore not archived as frequently.
When comparing the memento counts shown in Table \ref{tab:news_sites_mcounts}, washingtonpost.com has over 3000 mementos for each month, while \textit{chicagotribune.com} has 400 to 600 mementos per month.
This is an indication of the popularity of these sites, showing that less well known websites are less likely to be archived as frequently.

Many of the mementos requested at 8 PM Eastern Time returned mementos close to the request time of 8 PM Eastern Time.
However, even if a site was archived thousands of times a month it does not mean that there would be a memento for every single hour or specified request time as indicated by \textit{npr.org} in Figure \ref{fig:boxplot}.
On average the time difference from the requested URI-M and the actual URI-M was 15.2 minutes, while the earliest memento collected was 286.5 minutes before the requested time and the latest memento collected was 532.4 minutes after the requested time.

\begin{figure}
\includegraphics[height=2.8in, width=3.3in]{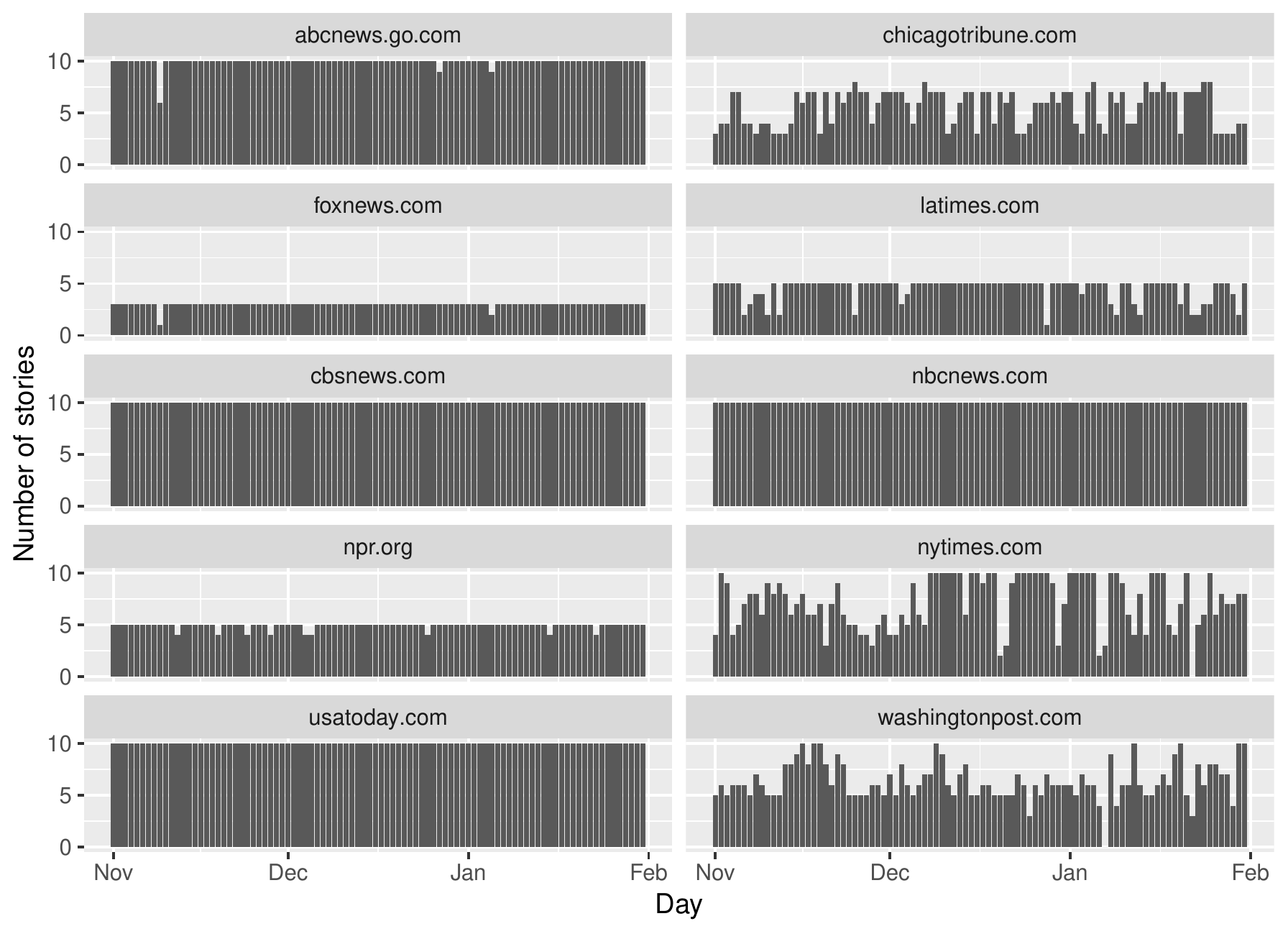}
\caption{Total number of stories for each website, with a limit of $k$ = 10, for each of the days from November 2016 to January 2017.}
\label{fig:total_stories}
\end{figure}

\subsection{Homepage parsing}

\begin{figure*}%if images are small remove * from figure
    \qquad
    \subfloat[NPR homepage]{{ \includegraphics[width=0.44\textwidth]{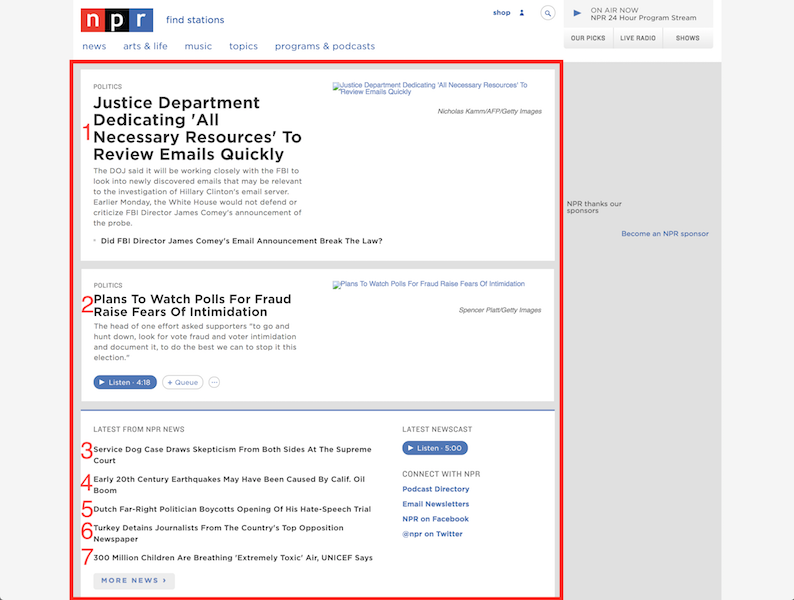} }}
    \qquad
    \subfloat[USA Today homepage]{{ \includegraphics[width=0.44\textwidth]{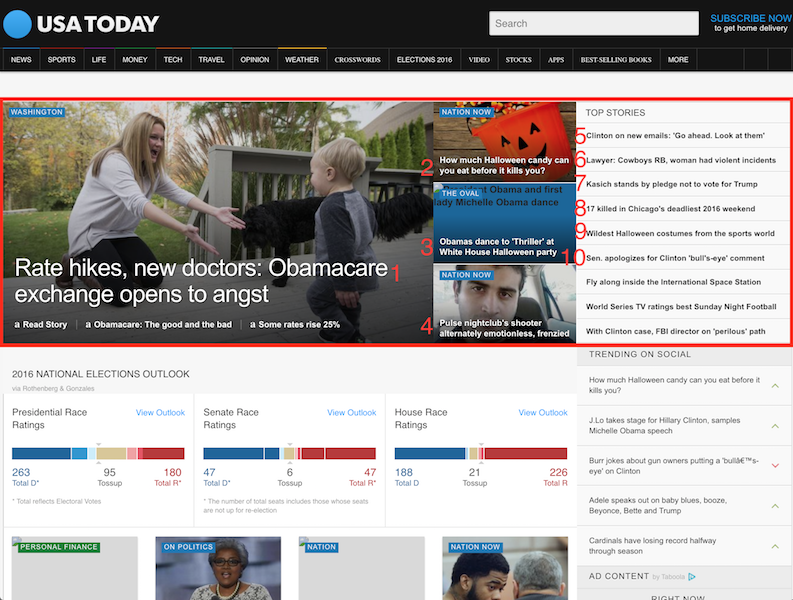} }}
    \qquad
    \subfloat[New York Times homepage]{{ \includegraphics[width=0.44\textwidth]{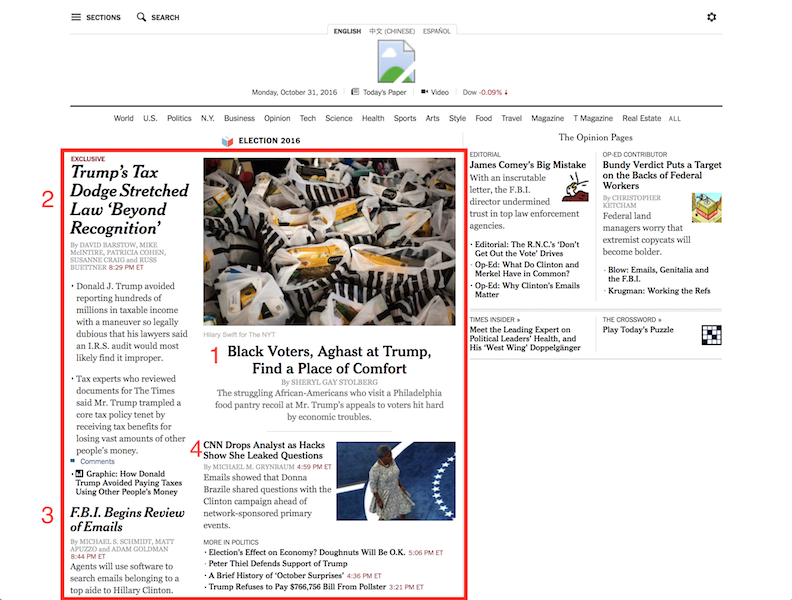} }}
    \qquad
    \subfloat[ABC News homepage]{{ \includegraphics[width=0.44\textwidth]{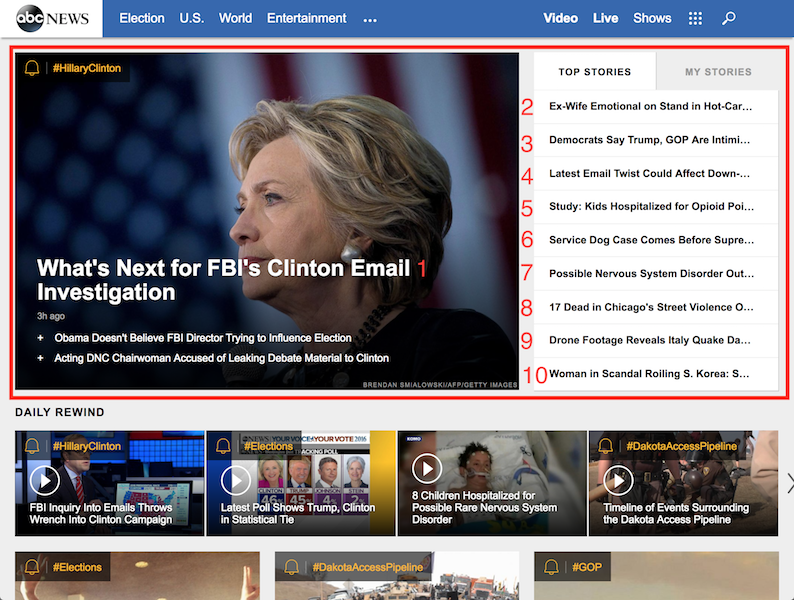} }}
    \qquad	
    \caption{Three mementos of news sites taken from November 1, 2016 1AM GMT (October 31, 2016 8PM Eastern Time). The number indicate the order of the stories processed and highlights the news site stories selected using CSS selectors. Number 1 indicates the Hero story for each news site. This day had cosine similarity score of 0.193 for $k$ = 3 stories from each news site.} 
    \label{fig:css_selectors}
\end{figure*}

Parsing titles and retrieving all the links from an HTML document is a relatively simple task.
This can be achieved by searching an HTML document for all the \textbf{<a>} elements that provide attributes, such as \textbf{href} or \textbf{src}.
However, news websites may often contain hundreds of links either to recommendations, opinion stories, or any other category labeled by the news website.
We use the term Hero stories in this paper to describe a story where the text is exaggerated across the top of a webpage or there is giant image indicating the main story at the current time on a website. For example, ``Rate hikes, new doctors: Obamacare exchange opens to angst'' as shown in Figure \ref{fig:css_selectors}b.
Our target stories were centered around the Hero stories, where $k = 1$, or the main headlines. 
This often appears at the top of a news website and is usually presented when the page finishes loading without the need to scroll down.

To accomplish the task of custom parsing, we created a parser\footnote{https://github.com/oduwsdl/top-news-selectors} utilizing Cascading Style Sheet (CSS) selectors, which are used to target HTML elements on web resources \cite{Selector3}.
Our parser utilizes the Python package Beautiful Soup\footnote{https://pypi.python.org/pypi/beautifulsoup4} to access CSS selectors inside HTML documents.
Although CSS selectors are often used to apply formatting for a webpage,  they also have applications to retrieve an element on a webpage.
Therefore, we can use CSS selectors to filter elements based on their attributes, retrieve the attributes inside the elements, retrieve the children of elements of a specified selector, and retrieve the text based on a specified selector. 
CSS selectors prove to be fairly successful when testing against news websites, but when websites depend on HTML \textbf{iframes} or JavaScript-injected HTML, CSS selectors cannot be used to select content before it is loaded.

Identifying the stories to be considered as the Hero stories was done empirically depending on the layout of the webpage.
For example, if the stories seemed to appear in a single column on the webpage (Figure 4a), the Hero story was usually denoted by the top story on the webpage that contained an enlarged image or enlarged text.
Other websites try to emulate a newspaper's format by applying multiple columns in a single view, showing different categories of news available on their website.
We found that when choosing the Hero story for this type of format the main stories were often in the central column (Figure \ref{fig:css_selectors}c), while the far right column stories were often opinion based or generated in real time.

Many of the websites explored in this paper self-identify the top stories presented in their website's HTML representation.
For example, USA Today in Figure \ref{fig:css_selectors}b shows the section our parser selects and the Hero story is identified by the selector ``a.hfwmm-primary-hed-link''. There is only one occurrence of this selector in the document.
News sites that did not label their content with an obvious CSS selector were parsed by taking the top stories from the leftmost and center stories.

After identifying all of the stories for the new sites using CSS selectors we performed requests to the Internet Archives Memento API \cite{RFC7089} to retrieve the content of each of the stories as provided from each memento.
Figure \ref{fig:total_stories} indicates the number of stories we identified for each news site for over the course of three months.
We found that there were 25 HTTP 404 response codes, identified as archived paywalls, and two ``infinite'' redirection loops from the unique story URIs collected.
For all documents that returned a 200 HTTP response code, we processed their HTML by extracting the text content through boilerplate removal with Python-Boilerplate\footnote{https://github.com/misja/python-boilerpipe} which we have found to consistently provide the best results \cite{boilerplateRemoval}.

\subsection{Similarity Metrics} \label{sim_metrics}

The collection similarity score, $s \in [0, 1]$, is a single value which quantifies the degree of similarity within the documents in a collection. 
A similarity score of 0 means all documents in the collection have no vocabulary in common, while a similarity score of 1 indicates maximum similarity (duplicate content). 
Given a collection of URIs, $C, (|C| = n$), the collection similarity score was calculated as follows:

\begin{enumerate}%[wide, labelwidth=!, labelindent=0pt]
  \item \textbf{Representation:} All the documents in the collection were represented as vectors. 
  Specifically, the vector representation is where each document $d_i$  in the collection $C$ is represented as a vector of TF-IDF values.
  
%  \begin{enumerate}%[wide, labelwidth=!, labelindent=0pt]
%    \item \textit{Vector representation:} each document, $d_i$ in the collection, $C$ was represented as a vector of TF-IDF values,
%    \item \textit{Set representation:} each document, $d_i$ in the collection, $C$ was represented as a set of entities. Entities consist of proper nouns of persons, locations, organizations, dates, times, percent, and miscellaneous. The entities were extracted using the Stanford Named Entities Recognition System \cite{finkel2005incorporating}.
%  \end{enumerate}
  
  \item \textbf{Pairwise similarity:} The pairwise similarity of all the documents were calculated to populate a similarity matrix, $\boldsymbol{D} \in \mathbb{R}^{n \times n}$. 
%  The similarity between a pair of documents was calculated using \textit{Vector representation similarity}. 
  Given a pair of documents, $d_i$ and $d_j$ in vector space, the similarity between the documents was calculated using the cosine similarity metric.
%  \begin{enumerate}
%    \item \textit{Vector representation similarity:} given a pair of documents, $d_i$ and $d_j$ in vector space, the similarity between the documents was calculated using the cosine similarity metric.
%    \item \textit{Set representation similarity:} given a pair of documents, $d_i$ and $d_j$ in set representation, the similarity between the pair of documents, $sim(d_i, d_j)$ was calculated using a weighted Jaccard-Overlap similarity method: 

%    \begin{equation}
%      sim(d_i, d_j) = 
%      \begin{cases}
%           1 & \text{; if $\alpha.J(d_i, d_j) + (1 - \alpha).O(d_i, d_j) \geq 0.27$} \\
%           0 & \text{; otherwise} \\
%      \end{cases}
%    \end{equation}
%    
%    $J(d_i, d_j)$ is the Jaccard index of both documents, $J(d_i, d_j) = \frac{|d_i \cap d_j|}{|d_i \cup d_j|}$, and $O(d_i, d_j)$, is the Overlap coefficient of both documents, $O(d_i, d_j) = \frac{|d_i \cap d_j|}{min(d_i, d_j)}$.
%  \end{enumerate}
  
  \item \textbf{Collection similarity score:} Given an \textit{all-ones matrix}, $\boldsymbol{O} \in \mathbb{R}^{n \times n}$, and a square matrix, $\boldsymbol{N} \in \mathbb{R}^{n \times n}$, with zeros on the main diagonal and ones everywhere else we can calculate the collection similarity $s$.
  For example if $\boldsymbol{N} \in \mathbb{R}^{3 \times 3}$,
  \[
    \boldsymbol{N} =
    \begin{bmatrix}
      0 & 1 & 1 \\
      1 & 0 & 1 \\
      1 & 1 & 0
    \end{bmatrix}\\ \\
   \]
   The collection similarity, $s$ would be calculated as follows:
  $s = \frac{||\boldsymbol{N.D}||_F}{||\boldsymbol{N.O}||_F}$
  , where $||\boldsymbol{A}||_F$ is the Frobenius norm:\\
  
 \(
  ||\boldsymbol{A}||_F = \sqrt{\sum_{i=1}^{m}\sum_{j=1}^{n}|a_{i,j}|^2}
 \)
\end{enumerate}

\section{Experiment Results}

\begin{figure*}%if images are small remove * from figure
    \includegraphics[width=0.98\textwidth]{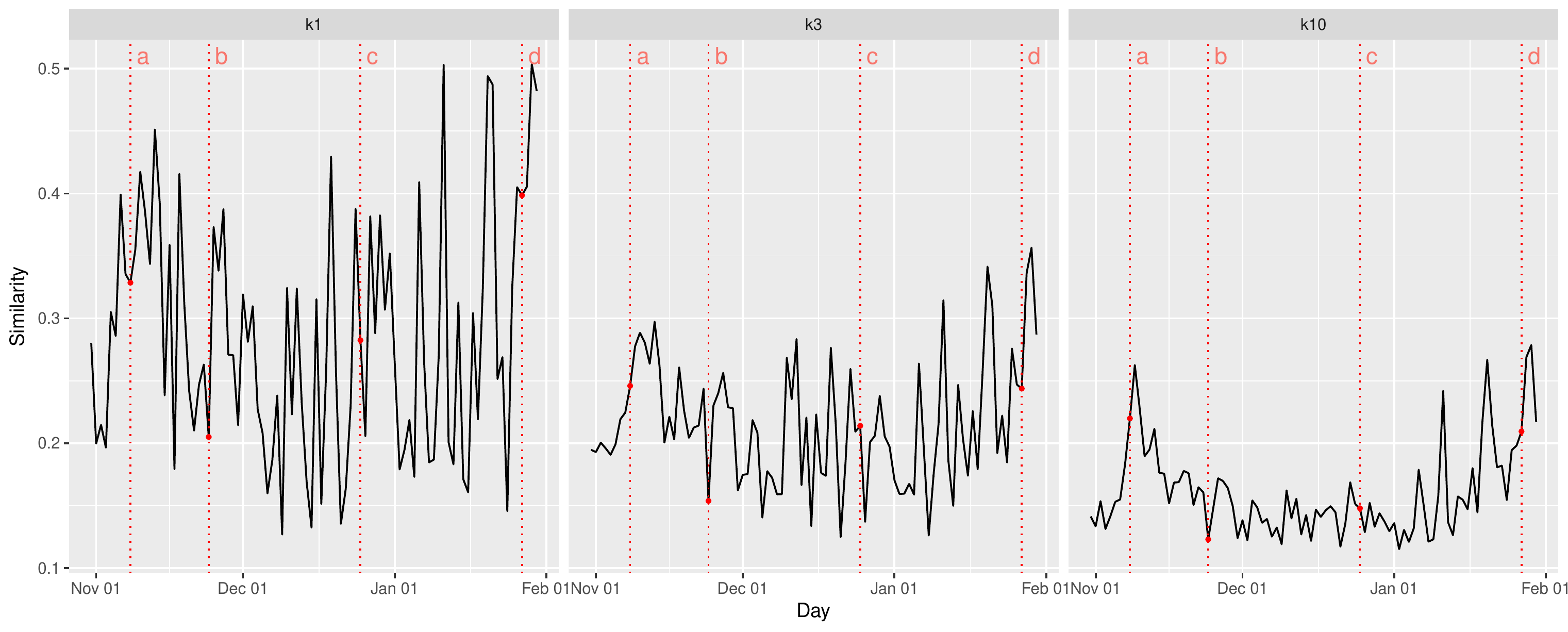}
    \caption{Cosine similarity scores for a given each day where  $k = {1, 3, 10}$  are the maximum number of stories for each news organization. Each graph is labeled with points of well known events: (a) Election Day (November 8, 2016), (b) Thanksgiving Day (November 24, 2016), (c) Christmas Day (December 25, 2016), and (d) Travel Ban (Executive Order 13769) comes into effect (January 27, 2017). }
    \label{fig:cos_sim_line}
\end{figure*}

Using the mementos collected for the ten news sites, we used our parser to extract each of the Hero stories and the subsequent headlines providing us with the text and the URI.
For the three months we extracted a total of 2349 unique memento URIs\footnote{We make our dataset publicly available at: \\https://github.com/grantat/news-similarity}. 
We found that some websites used their CSS selectors for identifying structure and order priority of content. 
For example, \textit{chicagotribune.com} had a Hero selector named ``h2.trb\_outfit\_primaryItem\_article\_title'' and a selector for identifying a section for leading headlines named ``.trb\_outfit\_list\_headline\_a''.
Some of the websites, for example \textit{cbsnews.com}, did not self identify where their headlines should cutoff. 
This would for some days lead to upwards of identifying twenty stories while other news sites would identify in a range of three to ten stories.
An example of this is shown in Figure \ref{fig:css_selectors}, where \textit{USA Today} has at least ten top stories found while we identify seven and four stories for \textit{NPR} and \textit{New York Times}, respectively.
Due to this reason, we decided to evaluate similarity limiting the number stories to a maximum of $k = {1, 3, 10}$ for each news site.
This means, when $k$ = 1 we would have a maximum of ten stories, one story from each news site homepage. 
When $k$ = 3 we would have a maximum of thirty stories, a maximum of three stories from each news site homepage.
When $k$ = 10 we would have a maximum of one hundred stories, a maximum of ten stories from each news site homepage.
We discovered that there were occasions when accessing the Internet Archive Memento API, in which some of the stories requested either resulted in 404, 301, to many redirects, response codes.
Therefore we excluded such URIs from the similarity calculation.

The position of stories in the HTML representations of the news site determines the order in which stories are determined for relevance, i.e. the third story found is story number three when $k = 3$.
For the three mementos shown in Figure \ref{fig:css_selectors}, there are stories that by just looking at their titles we can determine there is a similarity between them. 
For example, \textit{NPR} recognizes their Hero story as ``Justice Department Dedicating `All Necessary Resources' To Review Emails Quickly'' while \textit{USA Today} and \textit{New York Times} recognize their 5th and 3rd positioned top stories similar to \textit{NPR}'s Hero story.
For this day, if we take only $k$ = 1 story for each news site, the similarity would be affected due to the recognition of this story's importance not being widely recognized as the Hero story for every news site.
However, if we move on to $k$ = 3 then the \textit{New York Times} will have included their coverage of this story but also including two other stories that may not have a high similarity.

When identifying significant events we observed that political events had the most influence on the similarity scores.
Shown in Figure \ref{fig:cos_sim_line}, November 8, 2018 Election Day is an easily identifiable sequence of events but the peak of the similarity for this event occurs days after Election Day.
When the executive order started on January 28, 2017 this period also shows that there is a buildup of events in similarity, but the peak of similarity scores is after the start date.
This signifies that there is a delay in synchronization among news websites until the events have reached a critical point.

Figure \ref{fig:cos_sim_line} shows the cosine similarity scores, when $k$ = ${1, 3, 10}$, from November 2016 to January 2017 highlighting four different well known events.
For each different $k$ value we found the order of similarity between the four outlined events from highest to lowest was: Executive Order 13769 start date, Election Day, Christmas day, and Thanksgiving day.
National holidays, such as Thanksgiving and Christmas, had relatively low similarity scores when $k$ = 10.
Specifically, Thanksgiving day was the 8th lowest similarity score of the 92 days compared.
This shows that there is wider variance of stories and a lower synchronicity across the ten U.S. news sites on national holidays.

\subsection{Top-k Stories} \label{topksect}

When we limited the number of stories to the top $k$ = 1 stories, we were essentially taking the Hero story from each of the news websites.
This meant that we had a maximum of 10 stories for each of the days tested.
We observed that when we limit the number of stories to $k = 1$ that cosine similarity becomes high for significant events but also has very high variance among days.
Figure \ref{fig:cos_sim_line} shows that for November 8, 2016 there is a significant event that occurred preempted by a buildup of news similarity. 
Within a few days of the event the cosine score begins to drop indicating the story has either become less relevant or there is just less synchronization among news sites.
Cosine score peaks during the end of January 2017 also preempted by a buildup of news similarity.
During November 24, Thanksgiving day, there is a slight rise in similarity, however it is shown that the importance of this event is negligible compared to other stories that occurred around this day.

For $k = 3$ stories, Figure \ref{fig:cos_sim_line} shows a decline in similarity due to the introduction of more stories. 
%This increased the maximum number of stories in the set evaluated overall to 30.
The lowest similarity was seen when we took a maximum of $k = 10$ stories.
This shows that as the number of stories increased the overall confidence that stories were similar decreased.
%Table \ref{tab:sim_scores} shows the minimum, maximum, and average similarity measures for the entire 30 day period. 
%It also shows the decrease in overall similarity where the entity measure showed a higher average similarity for each top-k stories

\begin{figure*}%if images are small remove * from figure
    \qquad
    \subfloat[Fox News homepage]{{ \includegraphics[width=0.42\textwidth]{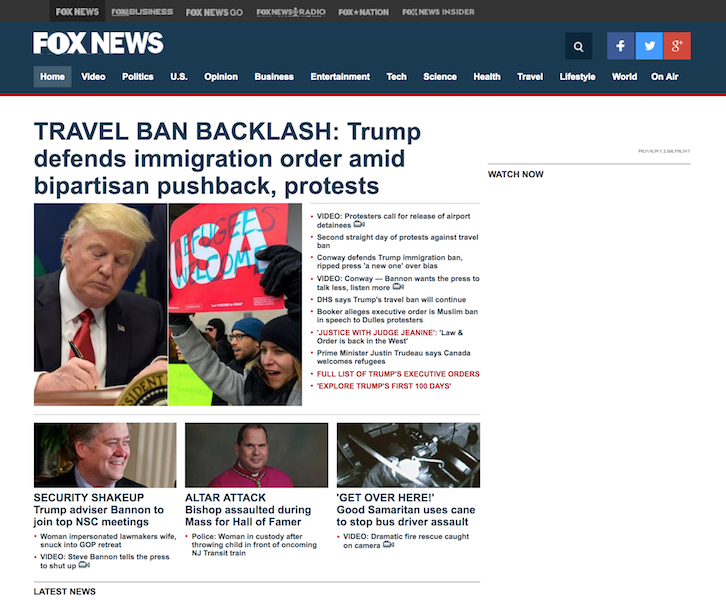} }}
    \qquad
    \subfloat[USA Today homepage]{{ \includegraphics[width=0.42\textwidth]{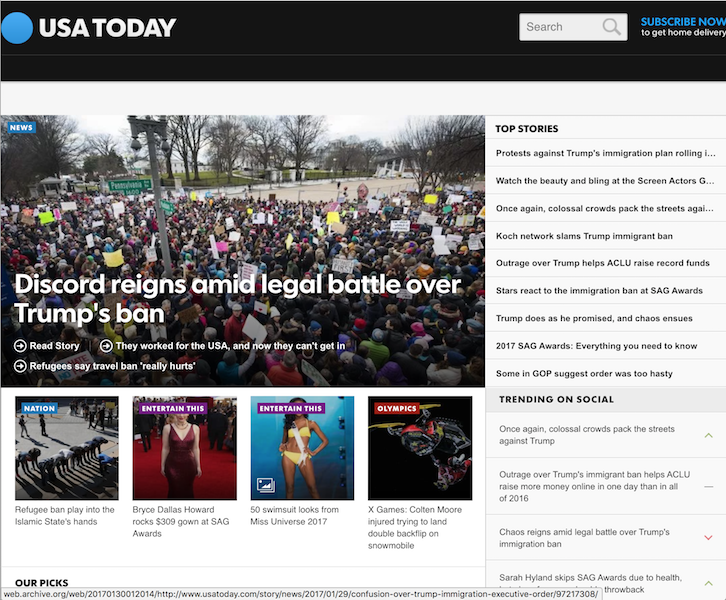} }}
    \qquad
    \subfloat[ABC News homepage]{{ \includegraphics[width=0.42\textwidth]{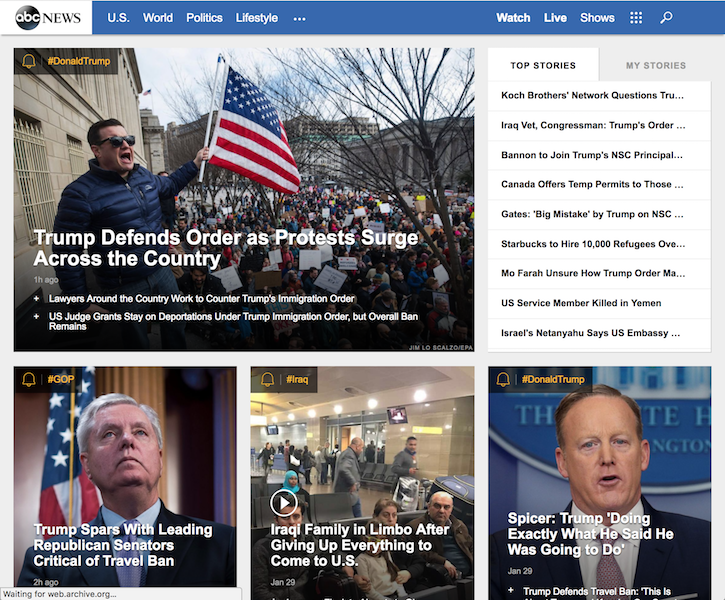} }}
    \qquad
    \subfloat[CBS News homepage]{{ \includegraphics[width=0.42\textwidth]{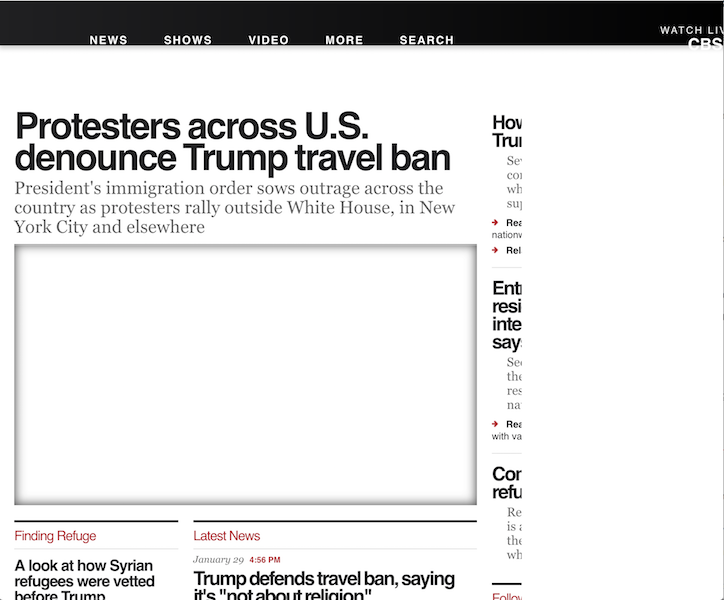} }}
    \qquad
    \caption{Four mementos of news sites taken from January 29, 2017 8PM Eastern Time. 
    This day had the highest cosine similarity score for each maximum $k$ value, where $k$ = 1 was 0.504, $k$ = 3 was 0.357, and $k$ = 10 was 0.279.
    These mementos do not discuss a national holiday and are a separate event from the presidential election but showed a high similarity because they address the same Hero story, $k$ = 1. 
    The Hero story for each news site was related to the ``Trump travel ban.''}
    \label{fig:nov11_sim}
\end{figure*}

The Election Day has a clear buildup of similarity scores.
Other days such as November 11, Veterans Day, also had a high similarity.
Figure \ref{fig:nov11_sim} shows there is a Hero story, related to the Travel Ban, shared across three different news sites.
This shows that using these similarity metrics that we are able to recognize these synchronous events.

% Table here

\begin{table}
\caption{Similarity score metrics for top-$k$ stories for the 3 selected months.}
\label{tab:sim_scores}
\begin{minipage}{\columnwidth}
\begin{center}
\begin{tabular}{l r | r r r}
\toprule
\textbf{Measure} & \textbf{Top-k} & \textbf{Min} & \textbf{Mean} & \textbf{Max} \\
\midrule
 	    & 1 & 0.1268  & 0.2858  & 0.5037\\
Cosine & 3 & 0.1248 & 0.2160  &  0.3566\\
	    & 10 & 0.1150 &  0.1608 &  0.2786\\
% \hline
% 	  & 1 & 0.1491 & 0.3977 & 0.7303\\
%Entity &  3 & 0.1661 & 0.2676 & 0.4097\\
%          & 10 & 0.1366 & 0.1937 & 0.3306 \\

\bottomrule
\end{tabular}
\end{center}
%\bigskip
%\footnotesize\emph{Source:} This is a table
%sourcenote. This is a table sourcenote. This is a table
%sourcenote.
%\emph{Note:} This is a table footnote.
\end{minipage}
\end{table}

%\begin{figure*}%if images are small remove * from figure
%    \qquad
%    \subfloat[NBC homepage]{{ \includegraphics[width=0.42\textwidth]{\rImgs/parsing/nbc_nov11.png} }}
%    \qquad
%    \subfloat[New York Times homepage]{{ \includegraphics[width=0.42\textwidth,height=0.38\textwidth]{\rImgs/parsing/nytimes_nov11.png} }}
%    \qquad
%    \subfloat[Chicago Tribune homepage]{{ \includegraphics[width=0.42\textwidth]{\rImgs/parsing/chicagotrib_nov11.png} }}
%    \qquad
%    \subfloat[Chicago Tribune homepage]{{ \includegraphics[width=0.42\textwidth]{\rImgs/parsing/chicagotrib_nov11.png} }}
%    \qquad
%    \caption{Three mementos of news sites taken from November 11, 2016 8PM Eastern Time. These mementos for Veterans day don't discuss the national holiday and are a separate event from the presidential election but showed a high similarity because they address the same Hero story, k = 1. Their Hero story is related to ``Mike Pence becomes head of transition team''. The cosine similarity for this event day was 0.385 for k = 1. }
%    \label{fig:nov11_sim}
%\end{figure*}

\subsection{Differing Story Priority} 

\begin{figure*}%if images are small remove * from figure
    \qquad
    \subfloat[ABC News homepage]{{ \includegraphics[width=0.42\textwidth]{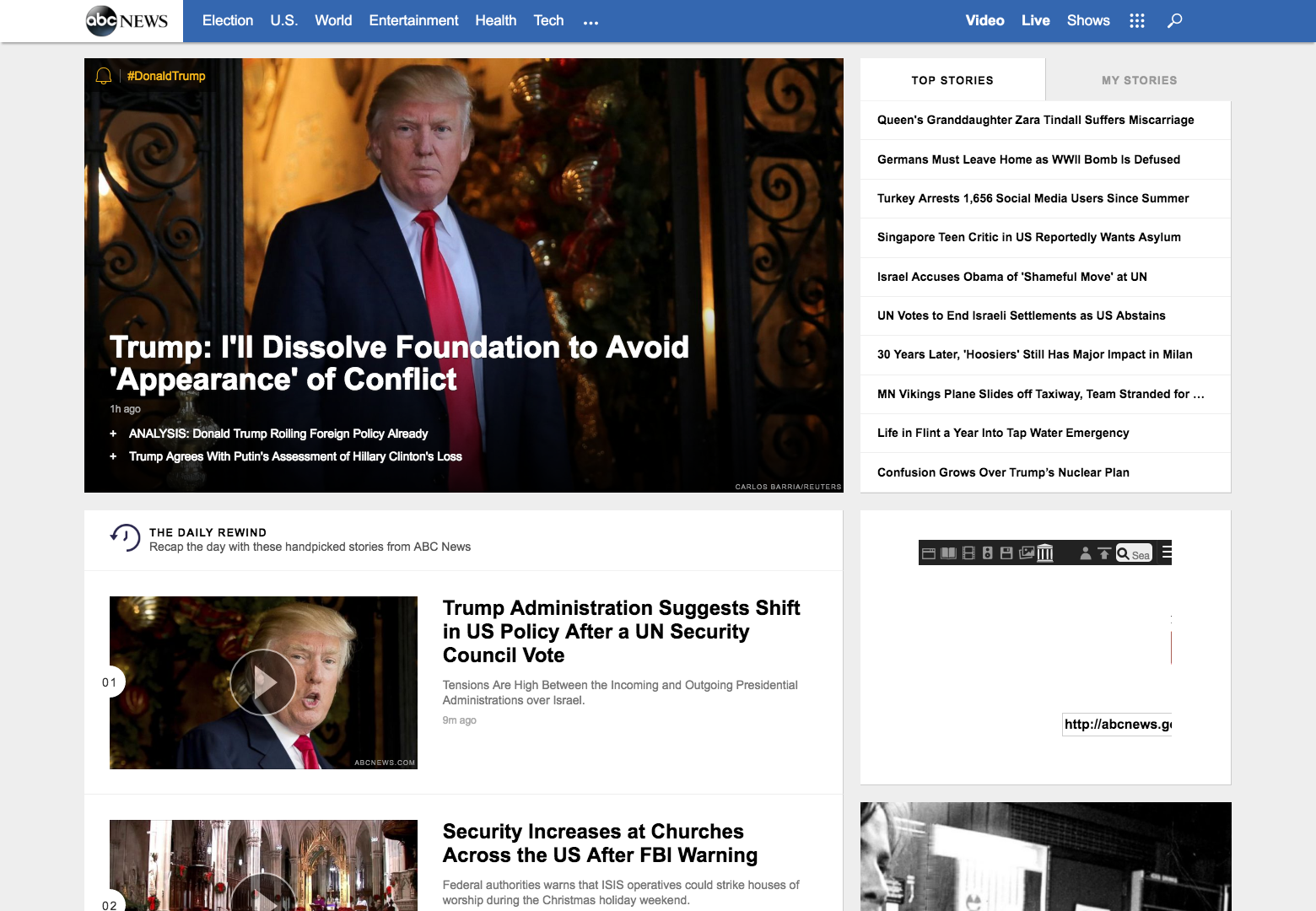} }}
    \qquad
    \subfloat[NBC News homepage]{{ \includegraphics[width=0.42\textwidth]{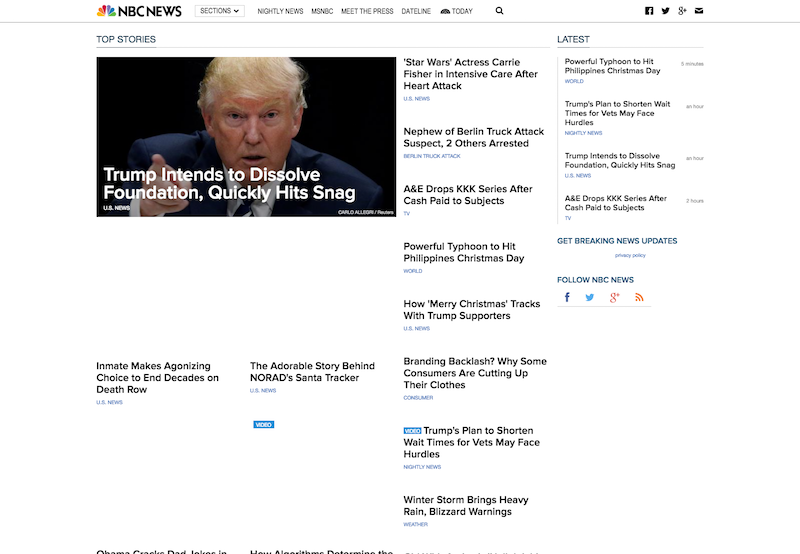} }}
    \qquad
    \subfloat[CBS News homepage]{{ \includegraphics[width=0.42\textwidth]{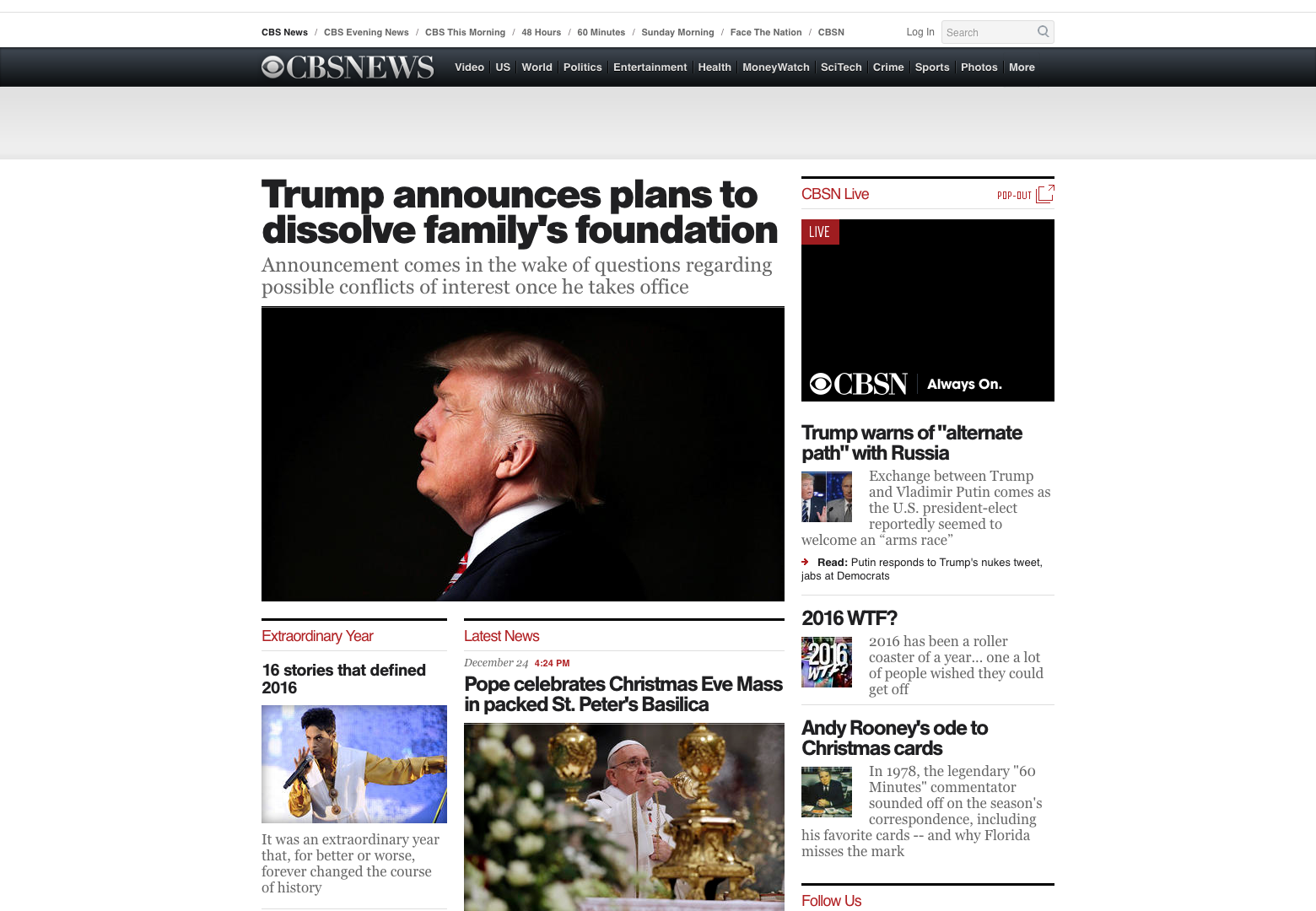} }}
    \qquad
    \subfloat[NPR News homepage]{{ \includegraphics[width=0.42\textwidth]{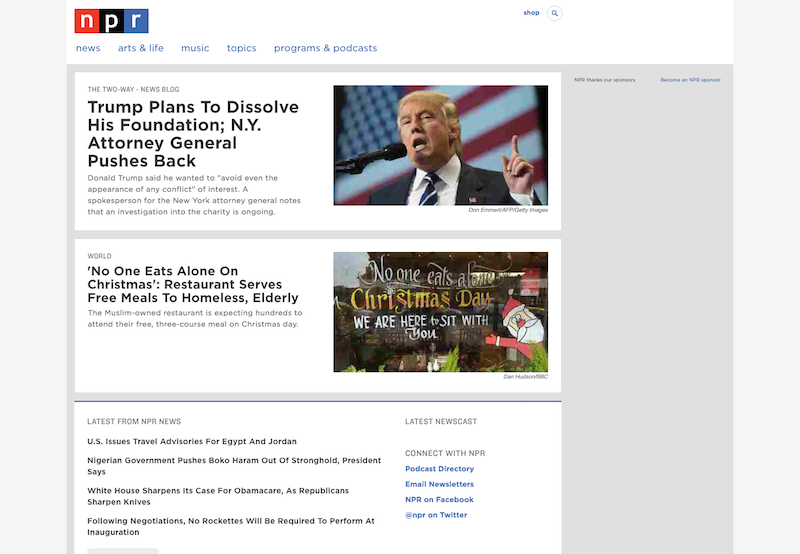} }}
    \qquad
    \subfloat[Fox News homepage]{{ \includegraphics[width=0.42\textwidth]{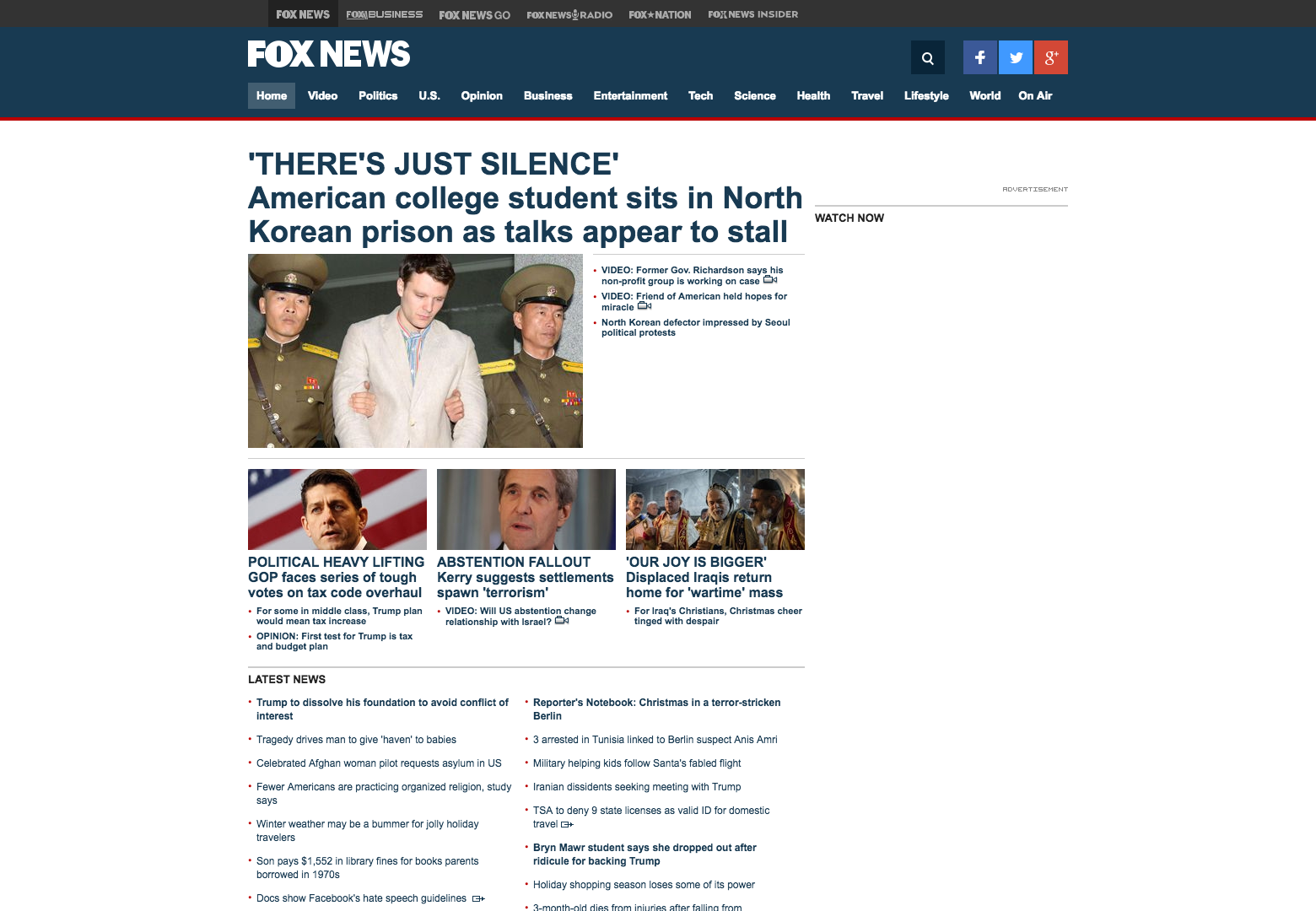} }}
    \qquad
    \subfloat[USA Today homepage]{{ \includegraphics[width=0.42\textwidth]{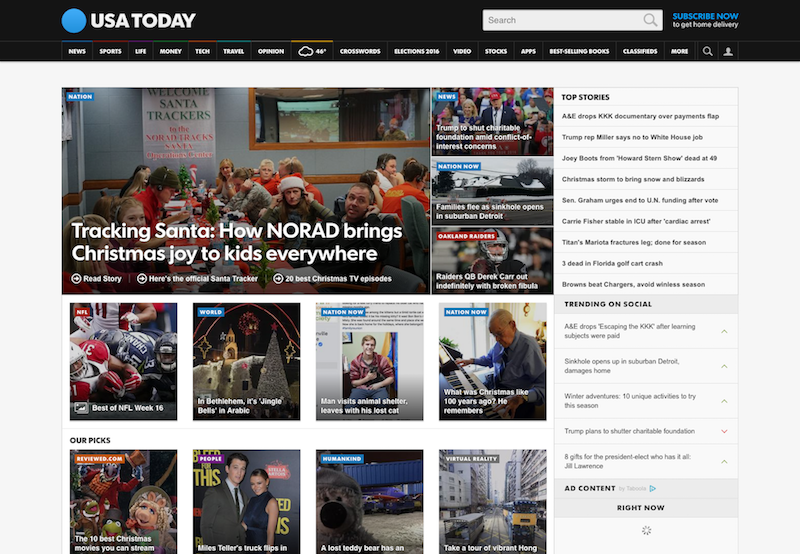} }}
    \qquad
    \caption{Six mementos of news sites taken from December 24, 2017 8PM Eastern Time. 
	Four of the news sites, shown in (a), (b), (c), and (d), have their Hero story relating to the topic ``Dissolve Trump Foundation.''
	Two of the news sites, shown in (e) and (f), have differing Hero stories as what they deem the more prominent story.
	}
    \label{fig:dec24figs}
\end{figure*}

There are many days where news homepages have homogeneous stories to prioritize as the Hero story.
For example, the travel ban that came into effect on January 27, 2017, shows the highest similarity score for Hero stories across each of the news sites.
This means that there was high synchronicity for each of the news sites' Hero stories.
Although this result is somewhat expected, what would happen if two of the ten news sites had a different Hero story based on a different topic?
This result would affect the similarity score for this day most likely resulting in a decline in similarity score.
An example of this is shown in Figure \ref{fig:dec24figs}, where on December 24, 2016 there were at least four news sites' Hero stories, including \textit{NPR}, \textit{NBC News}, \textit{ABC News}, and \textit{CBS News}, reporting on a political story about dissolving the Trump Foundation.
Two other news sites had Hero stories reporting on differing topics, one story about the approaching national holiday, Christmas, and the other story about another political matter.
It is worth noting that the Hero story from the first four news sites, shown in Figures \ref{fig:dec24figs}a - \ref{fig:dec24figs}d, is present on the two other news sites, Figures \ref{fig:dec24figs}e and \ref{fig:dec24figs}f, but is not deemed by the respective news sites to be the Hero story. 

News placement on a news site is important as it relates to the measuring news similarity, identifying news bias, and identifying story importance.
In regards to news similarity, if a story is presented as the Hero story for the majority of the sites but is considered the third story of other sites this could cause a decrease in similarity for $k = 1$ for the specified day.
However, if we were to increase the the number of stories taken from each site up to $k = 3$ stories it could potentially increase the similarity if the added stories from each site was the same topic of the original Hero story used by the majority of sites. 

\subsection{Election Day Influences} \label{el_day_influences}

\begin{figure*}%if images are small remove * from figure
    \qquad
    \subfloat[USA Today homepage]{{ \includegraphics[width=0.4\textwidth, height=0.35\textwidth]{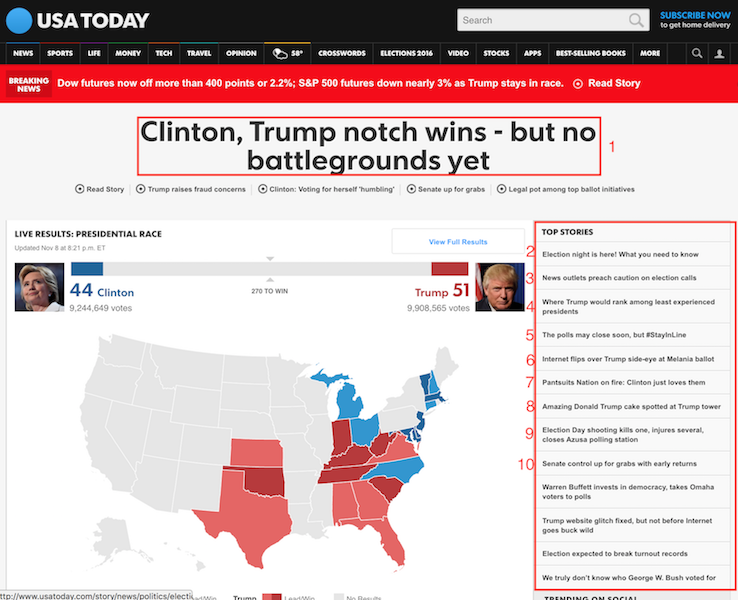} }}
    \qquad
    \subfloat[New York Times homepage]{{ \includegraphics[width=0.4\textwidth]{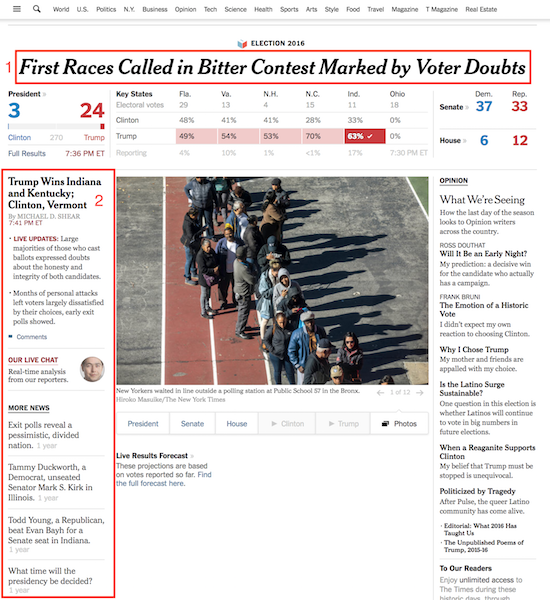} }}
    \qquad
    \subfloat[Los Angeles Times homepage]{{ \includegraphics[width=0.4\textwidth]{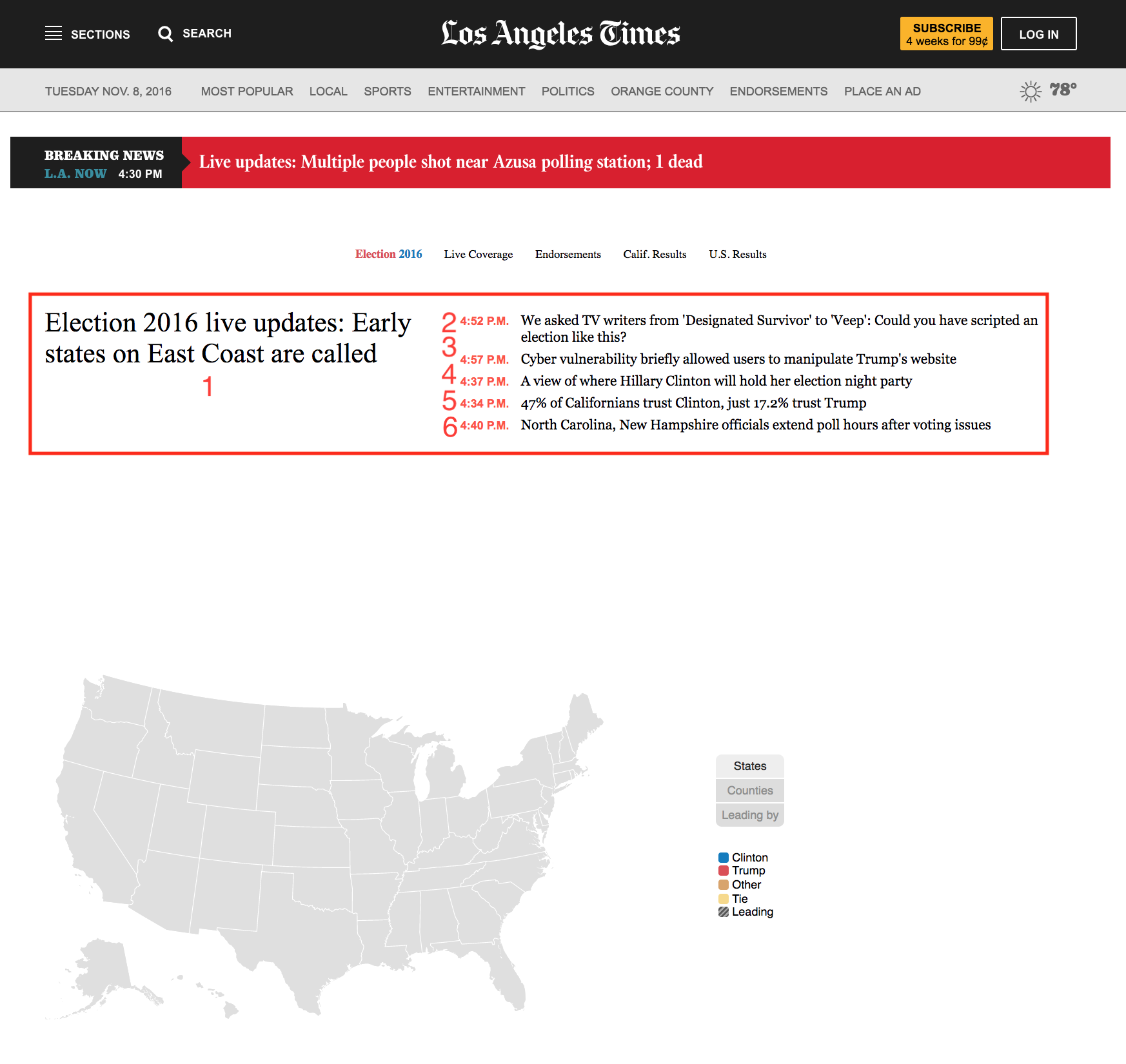} }}
    \qquad
    \caption{News sites format updated to track Election Day (November 8, 2016) progress for candidates: (a) USA Today introduces a United States map tracking candidate progress altering the naming conventions of their CSS selectors, (b) New York Times introduces a table of percentages of electoral votes, and (c) Los Angeles Times also introduces a world map to track candidate progress.}
    \label{fig:elect_day_docs}
\end{figure*}

As shown in Figure \ref{fig:cos_sim_line}, the election is a noticeable event due to high similarity scores.
However these news sites also make this event noticeable by providing a new layout.
When we first constructed our parser we noticed that using only a single set of CSS selectors proved to be ineffective when trying to parse headlines during the election.
Between November 7th - 11th, we found that five of the ten news sites altered their document representation just for the United States presidential election.
When these sites updated their document representations, some of them completely changed the naming conventions of CSS selectors but some of them still kept their previous selectors with this updated document.
We therefore prioritized Election Day selectors over the default selectors.
This also included the introduction of new naming conventions for HTML element \textbf{class} and \textbf{id} attributes.
To address this we added multiple selectors for these websites, some for both the Hero stories and also the subsequent headlines. 

These updates to their site layout meant that these sites recognized the significance of this event and chose to provide a new document representations to emphasize the importance of this event.
Figure \ref{fig:elect_day_docs} shows these changes and that many sites often included a United States map to track the progress of a state's electoral college votes for either presidential candidate.
After November 11th these sites returned to their original HTML representation.
%The highlighted portions of Figure \ref{fig:total_stories} indicate the days in which document representations started to change as well as an overlap of the election period.

%\begin{figure}
%\includegraphics[height=2.8in, width=3.3in]{\rImgs/k3/cos_sim_line_k3.pdf}
%\caption{Cosine and Entity Similarity per day for k = 3}
%\label{fig:k3_sim}
%\end{figure}
%
%\begin{figure}
%\includegraphics[height=2.8in, width=3.3in]{\rImgs/k10/Col_sim.pdf}
%\caption{Cosine and Entity Similarity per day for k = 10}
%\label{fig:k10_sim}
%\end{figure}

\section{Future Work}

For our experiment we only took news homepage mementos for a period of three months, we would like to extend the range to see how similarity changes.
We selected a single time in day to be used repeatedly across our time period, much like Klein and Broadwell \cite{Klein:2015}.
We would like to perform continuous observation with shorter intervals on a news site homepage and see how the document changes as well as the similarity to other news sites changes.
We realized that CSS selectors are an easily accessible format that allows us to select stories but as news websites update their document representations and change class naming conventions we also have to update the range of selectors our parser uses.

We found that a majority of titles for news homepage stories were actually shortened or summarized titles of the actual stories they reference.
We wish to analyze this further and see how the similarity of titles presented on the homepage differs from what is actually considered the true title of a story. 

\section{Conclusions}

The preservation of news is a valuable part of saving the memory of important historical events. 
Archived news pages provide a valuable opportunity for studying and analyzing events in a manner not possible on the live web. 
We provide tools to aid the analysis of archived news webpages in this work by introducing tools for parsing select HTML news sites for Hero and headline stories using CSS selectors.
We explored measuring similarity for ten U.S. news sites using the cosine similarity measure. 
We also discuss how news sites may alter their document representations for significant events such as a presidential election.
%Its shown that using either of the similarity measures for a news collection can be used to identify significant events.
%One of the goals of identifying a specific day where the similarity can be higher than others, however this could not be identified using a single measurement for a given day.
We define a method of mining web archives, specifically mining archived news.
We identify potential hazards when choosing mementos and describe the choices for which archived news sources are applicable for experimenting upon.

Our experiments over a three month period have shown that as the number of stories increase, the overall similarity decreased.
%Using these similarity metrics we identified the entity method to have the highest overall average similarity regardless of the value of k.
Using the calculated cosine scores we identify a decline, from 0.417 to 0.343, in similarity after the U.S. election period has passed, which indicates news sites pursuing other stories and decreased synchronization.
Our results show that we can identify synchronous stories besides related national events for a given day. 
This enables identifying the rise and decline of coverage using similarity.

%\end{document}  % This is where a 'short' article might terminate

%Of course, reading the source code is always useful.  The file
%\path{acmart.pdf} contains both the user guide and the commented
%code.

\begin{acks}
This work supported in part by NSF III 1526700 and IMLS LG-71-15-0077-15.
\end{acks}

%
% End content
%

%\input{samplebody-conf}

\bibliographystyle{ACM-Reference-Format}
\bibliography{gatkins_IPRES2018_arxiv}


\begin{thebibliography}{28}

%%% the Web address.

\ifx \showCODEN    \undefined \def \showCODEN     #1{\unskip}     \fi
\ifx \showDOI      \undefined \def \showDOI       #1{#1}\fi
\ifx \showISBNx    \undefined \def \showISBNx     #1{\unskip}     \fi
\ifx \showISBNxiii \undefined \def \showISBNxiii  #1{\unskip}     \fi
\ifx \showISSN     \undefined \def \showISSN      #1{\unskip}     \fi
\ifx \showLCCN     \undefined \def \showLCCN      #1{\unskip}     \fi
\ifx \shownote     \undefined \def \shownote      #1{#1}          \fi
\ifx \showarticletitle \undefined \def \showarticletitle #1{#1}   \fi
\ifx \showURL      \undefined \def \showURL       {\relax}        \fi
% invisible to TeX
\providecommand\bibfield[2]{#2}
\providecommand\bibinfo[2]{#2}
\providecommand\natexlab[1]{#1}
\providecommand\showeprint[2][]{arXiv:#2}

\bibitem[\protect\citeauthoryear{??}{exe}{2017}]%
        {execorder2017}
 \bibinfo{year}{2017}\natexlab{}.
\newblock \bibinfo{title}{{Executive Order Protecting the Nation from Foreign
  Terrorist Entry into the United States (Executive Order 13769)}}.
\newblock
  \bibinfo{howpublished}{\url{https://www.whitehouse.gov/presidential-actions/executive-order-protecting-nation-foreign-terrorist-entry-united-states/}}.
    (\bibinfo{year}{2017}).
\newblock


\bibitem[\protect\citeauthoryear{{Alexander Nwala}}{{Alexander Nwala}}{2017}]%
        {boilerplateRemoval}
\bibfield{author}{\bibinfo{person}{{Alexander Nwala}}.}
  \bibinfo{year}{2017}\natexlab{}.
\newblock \bibinfo{title}{{A survey of 5 boilerplate removal methods}}.
\newblock
  \bibinfo{howpublished}{\url{http://ws-dl.blogspot.com/2017/03/2017-03-20-survey-of-5-boilerplate.html}}.
    (\bibinfo{year}{2017}).
\newblock


\bibitem[\protect\citeauthoryear{Allan, Carbonell, Doddington, Yamron, and
  Yang}{Allan et~al\mbox{.}}{1998}]%
        {allan1998topic}
\bibfield{author}{\bibinfo{person}{James Allan}, \bibinfo{person}{Jaime~G
  Carbonell}, \bibinfo{person}{George Doddington}, \bibinfo{person}{Jonathan
  Yamron}, {and} \bibinfo{person}{Yiming Yang}.}
  \bibinfo{year}{1998}\natexlab{}.
\newblock \showarticletitle{Topic detection and tracking pilot study final
  report}.
\newblock  (\bibinfo{year}{1998}).
\newblock


\bibitem[\protect\citeauthoryear{{Alvin Chang}}{{Alvin Chang}}{2018}]%
        {voxFoxNewsVis}
\bibfield{author}{\bibinfo{person}{{Alvin Chang}}.}
  \bibinfo{year}{2018}\natexlab{}.
\newblock \bibinfo{title}{{The stories Fox News covers obsessively - and those
  it ignores - in charts}}.
\newblock
  \bibinfo{howpublished}{\url{https://www.vox.com/2018/5/30/17380096/fox-news-alternate-reality-charts}}.
    (\bibinfo{year}{2018}).
\newblock


\bibitem[\protect\citeauthoryear{Aturban, Kelly, Alam, Berlin, Nelson, and
  Weigle}{Aturban et~al\mbox{.}}{2018}]%
        {archivenow2018}
\bibfield{author}{\bibinfo{person}{Mohamed Aturban}, \bibinfo{person}{Mat
  Kelly}, \bibinfo{person}{Sawood Alam}, \bibinfo{person}{John Berlin},
  \bibinfo{person}{Michael Nelson}, {and} \bibinfo{person}{Michele Weigle}.}
  \bibinfo{year}{2018}\natexlab{}.
\newblock \showarticletitle{{ArchiveNow: Simplified, Extensible, Multi-Archive
  Preservation}}. In \bibinfo{booktitle}{\emph{Joint Conference on Digital
  Libraries (JCDL 2018)}}. \bibinfo{pages}{321--322}.
\newblock


\bibitem[\protect\citeauthoryear{Bergmark}{Bergmark}{2002}]%
        {bergmark2002collection}
\bibfield{author}{\bibinfo{person}{Donna Bergmark}.}
  \bibinfo{year}{2002}\natexlab{}.
\newblock \showarticletitle{Collection synthesis}. In
  \bibinfo{booktitle}{\emph{{Joint Conference on Digital Libraries (JCDL
  2002)}}}. \bibinfo{pages}{253--262}.
\newblock


\bibitem[\protect\citeauthoryear{Berlin}{Berlin}{2017}]%
        {Berlin:cnn}
\bibfield{author}{\bibinfo{person}{John Berlin}.}
  \bibinfo{year}{2017}\natexlab{}.
\newblock \bibinfo{title}{{CNN.com} has been unarchivable since {November} 1st,
  2016}.
\newblock
  \bibinfo{howpublished}{\url{http://ws-dl.blogspot.com/2017/01/2017-01-20-cnncom-has-been-unarchivable.html}}.
    (\bibinfo{date}{Jan.} \bibinfo{year}{2017}).
\newblock


\bibitem[\protect\citeauthoryear{Brunelle, Kelly, SalahEldeen, Weigle, and
  Nelson}{Brunelle et~al\mbox{.}}{2015}]%
        {brunelle2015not}
\bibfield{author}{\bibinfo{person}{Justin~F Brunelle}, \bibinfo{person}{Mat
  Kelly}, \bibinfo{person}{Hany SalahEldeen}, \bibinfo{person}{Michele~C
  Weigle}, {and} \bibinfo{person}{Michael~L Nelson}.}
  \bibinfo{year}{2015}\natexlab{}.
\newblock \showarticletitle{Not all mementos are created equal: Measuring the
  impact of missing resources}.
\newblock \bibinfo{journal}{\emph{International Journal on Digital Libraries}}
  \bibinfo{volume}{16}, \bibinfo{number}{3-4} (\bibinfo{year}{2015}),
  \bibinfo{pages}{283--301}.
\newblock


\bibitem[\protect\citeauthoryear{Celik, Etemad, Glazman, Hickson, Linss, and
  Williams}{Celik et~al\mbox{.}}{2011}]%
        {Selector3}
\bibfield{author}{\bibinfo{person}{Tantek Celik}, \bibinfo{person}{Elika~J.
  Etemad}, \bibinfo{person}{Daniel Glazman}, \bibinfo{person}{Ian Hickson},
  \bibinfo{person}{Peter Linss}, {and} \bibinfo{person}{John Williams}.}
  \bibinfo{year}{2011}\natexlab{}.
\newblock \bibinfo{title}{Selectors Level 3}.
\newblock   (\bibinfo{date}{Sept.} \bibinfo{year}{2011}).
\newblock
\urldef\tempurl%
\url{https://www.w3.org/TR/css3-selectors/}
\showURL{%
Retrieved January 15, 2017 from \tempurl}


\bibitem[\protect\citeauthoryear{Chakrabarti, Van~den Berg, and
  Dom}{Chakrabarti et~al\mbox{.}}{1999}]%
        {chakrabarti1999focused}
\bibfield{author}{\bibinfo{person}{Soumen Chakrabarti}, \bibinfo{person}{Martin
  Van~den Berg}, {and} \bibinfo{person}{Byron Dom}.}
  \bibinfo{year}{1999}\natexlab{}.
\newblock \showarticletitle{Focused crawling: a new approach to topic-specific
  Web resource discovery}.
\newblock \bibinfo{journal}{\emph{Computer networks}} \bibinfo{volume}{31},
  \bibinfo{number}{11} (\bibinfo{year}{1999}), \bibinfo{pages}{1623--1640}.
\newblock


\bibitem[\protect\citeauthoryear{de~Sompel, Nelson, and Sanderson}{de~Sompel
  et~al\mbox{.}}{2013}]%
        {RFC7089}
\bibfield{author}{\bibinfo{person}{H.~Van de Sompel}, \bibinfo{person}{M.
  Nelson}, {and} \bibinfo{person}{R. Sanderson}.}
  \bibinfo{year}{2013}\natexlab{}.
\newblock \bibinfo{booktitle}{\emph{HTTP Framework for Time-Based Access to
  Resource States -- Memento}}.
\newblock \bibinfo{type}{RFC} 7089. \bibinfo{institution}{RFC Editor}.
\newblock
\showISSN{2070-1721}


\bibitem[\protect\citeauthoryear{{Dhrumil Mehta}}{{Dhrumil Mehta}}{2017}]%
        {fivethirtyEight}
\bibfield{author}{\bibinfo{person}{{Dhrumil Mehta}}.}
  \bibinfo{year}{2017}\natexlab{}.
\newblock \bibinfo{title}{{All The Cable News Networks Are Covering The 'Russia
  Story' - Just Not The Same One}}.
\newblock
  \bibinfo{howpublished}{\url{https://fivethirtyeight.com/features/all-the-cable-news-networks-are-covering-the-russia-story-just-not-the-same-one/}}.
    (\bibinfo{year}{2017}).
\newblock


\bibitem[\protect\citeauthoryear{Farag, Lee, and Fox}{Farag
  et~al\mbox{.}}{2018}]%
        {farag2017focused}
\bibfield{author}{\bibinfo{person}{Mohamed~MG Farag}, \bibinfo{person}{Sunshin
  Lee}, {and} \bibinfo{person}{Edward~A Fox}.} \bibinfo{year}{2018}\natexlab{}.
\newblock \showarticletitle{Focused crawler for events}.
\newblock \bibinfo{journal}{\emph{{International Journal on Digital Libraries
  (IJDL)}}} \bibinfo{volume}{19}, \bibinfo{number}{1} (\bibinfo{year}{2018}),
  \bibinfo{pages}{1--19}.
\newblock


\bibitem[\protect\citeauthoryear{Faris, Roberts, Etling, Bourassa, Zuckerman,
  and Benkler}{Faris et~al\mbox{.}}{2017}]%
        {faris2017partisanship}
\bibfield{author}{\bibinfo{person}{Robert Faris}, \bibinfo{person}{Hal
  Roberts}, \bibinfo{person}{Bruce Etling}, \bibinfo{person}{Nikki Bourassa},
  \bibinfo{person}{Ethan Zuckerman}, {and} \bibinfo{person}{Yochai Benkler}.}
  \bibinfo{year}{2017}\natexlab{}.
\newblock \showarticletitle{Partisanship, Propaganda, and Disinformation:
  Online Media and the 2016 US Presidential Election}.
\newblock  (\bibinfo{year}{2017}).
\newblock


\bibitem[\protect\citeauthoryear{{Grant Atkins}}{{Grant Atkins}}{2018}]%
        {paywallAtkins}
\bibfield{author}{\bibinfo{person}{{Grant Atkins}}.}
  \bibinfo{year}{2018}\natexlab{}.
\newblock \bibinfo{title}{{Paywalls in the Internet Archive}}.
\newblock
  \bibinfo{howpublished}{\url{http://ws-dl.blogspot.com/2018/03/2018-03-15-paywalls-in-internet-archive.html}}.
    (\bibinfo{year}{2018}).
\newblock


\bibitem[\protect\citeauthoryear{Hamborg, Meuschke, and Gipp}{Hamborg
  et~al\mbox{.}}{2017}]%
        {matrixBased}
\bibfield{author}{\bibinfo{person}{F. Hamborg}, \bibinfo{person}{N. Meuschke},
  {and} \bibinfo{person}{B. Gipp}.} \bibinfo{year}{2017}\natexlab{}.
\newblock \showarticletitle{{Matrix-Based News Aggregation: Exploring Different
  News Perspectives}}. In \bibinfo{booktitle}{\emph{{Joint Conference on
  Digital Libraries (JCDL 2017)}}}. IEEE, \bibinfo{pages}{1--10}.
\newblock


\bibitem[\protect\citeauthoryear{Hansen and Paul}{Hansen and Paul}{2017}]%
        {hansen2017futureproofing}
\bibfield{author}{\bibinfo{person}{Kathleen~A. Hansen} {and}
  \bibinfo{person}{Nora Paul}.} \bibinfo{year}{2017}\natexlab{}.
\newblock \bibinfo{booktitle}{\emph{Future-proofing the news : preserving the
  first draft of history}}.
\newblock \bibinfo{publisher}{Rowman \& Littlefield},
  \bibinfo{address}{Lanham}.
\newblock
\showISBNx{9781442267121}


\bibitem[\protect\citeauthoryear{He and Parker}{He and Parker}{2010}]%
        {he2010topic}
\bibfield{author}{\bibinfo{person}{Dan He} {and} \bibinfo{person}{D~Stott
  Parker}.} \bibinfo{year}{2010}\natexlab{}.
\newblock \showarticletitle{Topic dynamics: an alternative model of bursts in
  streams of topics}. In \bibinfo{booktitle}{\emph{Proceedings of the 16th ACM
  SIGKDD international conference on Knowledge discovery and data mining}}.
  ACM, \bibinfo{pages}{443--452}.
\newblock


\bibitem[\protect\citeauthoryear{Klein and Broadwell}{Klein and
  Broadwell}{2015}]%
        {Klein:2015}
\bibfield{author}{\bibinfo{person}{Martin Klein} {and} \bibinfo{person}{Peter
  Broadwell}.} \bibinfo{year}{2015}\natexlab{}.
\newblock \showarticletitle{Analyzing News Events in Non-Traditional Digital
  Library Collections}. In \bibinfo{booktitle}{\emph{Joint Conference on
  Digital Libraries (JCDL 2015)}}. \bibinfo{pages}{191--194}.
\newblock
\showISBNx{978-1-4503-3594-2}


\bibitem[\protect\citeauthoryear{Lau, Collier, and Baldwin}{Lau
  et~al\mbox{.}}{2012}]%
        {lau2012line}
\bibfield{author}{\bibinfo{person}{Jey~Han Lau}, \bibinfo{person}{Nigel
  Collier}, {and} \bibinfo{person}{Timothy Baldwin}.}
  \bibinfo{year}{2012}\natexlab{}.
\newblock \showarticletitle{On-line Trend Analysis with Topic Models: \#
  twitter Trends Detection Topic Model Online.}. In
  \bibinfo{booktitle}{\emph{COLING}}. \bibinfo{pages}{1519--1534}.
\newblock


\bibitem[\protect\citeauthoryear{Nwala, Weigle, Ziegler, Aizman, and
  Nelson}{Nwala et~al\mbox{.}}{2017}]%
        {nwala2017local}
\bibfield{author}{\bibinfo{person}{Alexander~C Nwala},
  \bibinfo{person}{Michele~C Weigle}, \bibinfo{person}{Adam~B Ziegler},
  \bibinfo{person}{Anastasia Aizman}, {and} \bibinfo{person}{Michael~L
  Nelson}.} \bibinfo{year}{2017}\natexlab{}.
\newblock \showarticletitle{Local Memory Project: Providing Tools to Build
  Collections of Stories for Local Events from Local Sources}. In
  \bibinfo{booktitle}{\emph{Joint Conference on Digital Libraries (JCDL
  2017)}}. IEEE, \bibinfo{pages}{1--10}.
\newblock


\bibitem[\protect\citeauthoryear{{Reynolds Journalism Institute}}{{Reynolds
  Journalism Institute}}{2017}]%
        {dodgingMemoryHole}
\bibfield{author}{\bibinfo{person}{{Reynolds Journalism Institute}}.}
  \bibinfo{year}{2017}\natexlab{}.
\newblock \bibinfo{title}{{Dodging the Memory Hole 2017}}.
\newblock
  \bibinfo{howpublished}{\url{https://www.rjionline.org/events/dodging-the-memory-hole-2017}}.
    (\bibinfo{year}{2017}).
\newblock


\bibitem[\protect\citeauthoryear{{Roger Sollenberger}}{{Roger
  Sollenberger}}{2017}]%
        {Sollenberger}
\bibfield{author}{\bibinfo{person}{{Roger Sollenberger}}.}
  \bibinfo{year}{2017}\natexlab{}.
\newblock \bibinfo{title}{{How the Trump-Russia Data Machine Games Google to
  Fool Americans}}.
\newblock
  \bibinfo{howpublished}{\url{https://www.pastemagazine.com/articles/2017/06/how-the-trump-russia-data-machine-games-google-to.html}}.
    (\bibinfo{year}{2017}).
\newblock


\bibitem[\protect\citeauthoryear{Rosenthal}{Rosenthal}{2017}]%
        {archivePaywalls}
\bibfield{author}{\bibinfo{person}{David Rosenthal}.}
  \bibinfo{year}{2017}\natexlab{}.
\newblock \bibinfo{title}{{Talk at Spring 2013 CNI}}.
\newblock
  \bibinfo{howpublished}{\url{https://blog.dshr.org/2013/04/talk-at-spring-2013-cni.html}}.
    (\bibinfo{date}{December} \bibinfo{year}{2017}).
\newblock


\bibitem[\protect\citeauthoryear{{Shan Wang}}{{Shan Wang}}{2018}]%
        {shanwangny}
\bibfield{author}{\bibinfo{person}{{Shan Wang}}.}
  \bibinfo{year}{2018}\natexlab{}.
\newblock \bibinfo{title}{{Here's how The New York Times is trying to preserve
  millions of old pages the way they were originally published}}.
\newblock \bibinfo{howpublished}{\url{http://www.niemanlab.org/2018/04/}}.
  (\bibinfo{year}{2018}).
\newblock


\bibitem[\protect\citeauthoryear{{Taylor Adams, Jessia Ma and Stuart A.
  Thompson}}{{Taylor Adams, Jessia Ma and Stuart A. Thompson}}{2017}]%
        {nytimesStudy}
\bibfield{author}{\bibinfo{person}{{Taylor Adams, Jessia Ma and Stuart A.
  Thompson}}.} \bibinfo{year}{2017}\natexlab{}.
\newblock \bibinfo{title}{{Trump Loves 'Fox \& Friends.' Here's Why.}}
\newblock
  \bibinfo{howpublished}{\url{https://www.nytimes.com/interactive/2017/11/01/opinion/How-Fox-News-Covered-the-Manafort-Indictment.html}}.
    (\bibinfo{year}{2017}).
\newblock


\bibitem[\protect\citeauthoryear{Wolff}{Wolff}{2018}]%
        {fireFuryBook}
\bibfield{author}{\bibinfo{person}{Michael Wolff}.}
  \bibinfo{year}{2018}\natexlab{}.
\newblock \bibinfo{booktitle}{\emph{Fire and Fury: Inside the Trump White
  House}}.
\newblock \bibinfo{publisher}{Henry Holt and Co.}
\newblock
\showISBNx{1250158060}


\bibitem[\protect\citeauthoryear{Woolley and Guilbeault}{Woolley and
  Guilbeault}{2017}]%
        {woolley2017computational}
\bibfield{author}{\bibinfo{person}{Samuel~C Woolley} {and}
  \bibinfo{person}{Douglas~R Guilbeault}.} \bibinfo{year}{2017}\natexlab{}.
\newblock \showarticletitle{Computational Propaganda in the United States of
  America: Manufacturing Consensus Online}.
\newblock \bibinfo{journal}{\emph{Computational Propgaganda Research Project}}
  (\bibinfo{year}{2017}), \bibinfo{pages}{22}.
\newblock


\end{thebibliography}
%\bibliography{paper-bib}

\end{document}